\documentclass[a4paper,11pt]{article}
\usepackage{jheppub} % for details on the use of the package, please see the JINST-author-manual
\usepackage{lineno}
\usepackage{tikz}
\usepackage{tikz-cd}
\usepackage{graphicx}

\usepackage{array}
\usepackage{tabularx}
\usepackage{makecell}
\usetikzlibrary{decorations.markings}

\def\tr{\mathop{\mathrm{tr}}\nolimits}

\title{\boldmath Symmetry, Symmetry Topological Field Theory and von Neumann Algebra}

\author[a]{Qiang Jia}
\author[b]{Jiahua Tian}
\affiliation[a]{Department of Physics, Korea Advanced Institute of Science and Technology,\\
Daejeon, 34141, Korea}
\affiliation[b]{School of Physics and Electronics, East China Normal University,\\
500 Dongchuan Road, Shanghai, 200241, China}

\emailAdd{qjia1993@kaist.ac.kr, jtian1905@gmail.com}

\abstract{We study the additivity and Haag duality of the von Neumann algebra of a quantum field theory $\mathcal{T}_\mathcal{F}$ with 0-form (and the dual $(d-2)$-form) (non)-invertible global symmetry $\mathcal{F}$. We analyze the symmetric (uncharged) sector von Neumann algebra of $\mathcal{T}_\mathcal{F}$ with the inclusion of bi-local and bi-twist operators in it. We establish the connection between the existence of these non-local operators in $\mathcal{T}_\mathcal{F}$ and certain properties of the Lagrangian algebra $\mathcal{L}$ of the extended operators in the corresponding symmetry topological field theory (SymTFT). We prove that additivity or Haag duality of the symmetric sector von Neumann algebra is violated when $\mathcal{L}$ satisfies specific criteria, thus generalizing the result of Shao, Sorce and Srivastava to arbitrary dimensions. We further demonstrate the SymTFT construction via concrete examples in two dimensions.}

\begin{document}
\maketitle
\flushbottom

\section{Introduction}
\label{sec:intro}

Locality has been the central guiding principle in the searching for the fundamental laws of Nature in modern physics since the era of Einstein. Together with the principle of Quantum Mechanics, it is one of the pillars of Quantum Field Theory~\cite{Weinberg:1995mt, Weinberg:1996kr}. Over the past several decades, physicists have made tremendous efforts to formalize and understand the foundations of QFT. One framework that incorporates locality deeply in its roots is the approach of algebraic QFT~\cite{Haag:1963dh, Haag:1996hvx}. Following the proposal of~\cite{Segal:1947}, a $\mathbb{C}^*$-algebra $\mathcal{A}(R)$, which is taken to be a von Neumann algebra of bounded operators on the Hilbert space $\mathcal{H}$, is attached to a general region $R$ of spacetime~\cite{Haag:1962}. This approach to QFT, and its modification in the presence of gravity, are particularly useful in recent developments on the algebra of observables in de Sitter space~\cite{Chandrasekaran:2022cip, Witten:2023qsv}. Nevertheless, in this work, we restrict ourselves to the study of QFT without gravity.

One of the most important features of a QFT is its global symmetry. The observations in the early days~\cite{Alford:1990fc, Alford:1991vr, Bucher:1991bc, Alford:1992yx, NUSSINOV2009977, Witten:AdSCFT_TFT, Freed:FluxUncertainty, Pantev:2005zs, Pantev:2005wj, Pantev:2005rh, Hellerman:2006zs, Gukov:2006jk, Aharony:ReadingLines4D, Kapustin:2013uxa, Kapustin:2014gua} were coined in the foundational work~\cite{Gaiotto:GenSymm} as \emph{generalized global symmetries}, whose various facets, including higher form symmetries, higher group symmetries, and non-invertible symmetries have been extensively studied in the past decade~\cite{Albertini:HigherFormMth, Apruzzi:SymTFT, Apruzzi:GlobalForm_2group, Apruzzi:2group_6D, Apruzzi:Higher_Form_6D, Apruzzi:Holography_1form_Confinement, Apruzzi:NonInvert_Holography, Apruzzi:AspectsSymTFT, Acharya:2023bth, Bashmakov:NonInvClassS, 
vanBeest:SymTFT3D, BenettiGenolini:2020doj, 
Bergman:GenSymmHoloABJM, Bhardwaj:Higher_form_5D6D, Bhardwaj:2Group_S, 
Bhardwaj:NonInvHigherCat, Bhardwaj:GenChargeI, Bhardwaj:GenChargeII, Bhardwaj:1formClassS, Bhardwaj:AnomalyDefect, Bhardwaj:UniversalNonInv, Bhardwaj:UnifyingConstructionNonInv, Bhardwaj:NonInvWeb, Braeger:2024jcj, Choi:NonInv3+1, Choi:2022jqy, Cvetic:0form_1form_2group, Cvetic:HigherFormAnomaly, Cvetic:Fluxbrane, 
Cvetic:GensymmGravity, Cvetic:2024dzu, Cvetic:2025kdn, DelZotto:2groupMth, DelZotto:HigherFormAD, DelZotto:HigherFormOrbifold, Etheredge:BraneSymm, GarciaEtxebarria:BraneNonInv, GarciaEtxebarria:Goldstone, Gukov:2020btk, Heckman:Branes_GenSymm, Heckman:2024oot, Hsieh:Inflow, Hubner:GenSymm_EFib, Jia:2025jmn, Kaidi:KW3+1D, Kaidi:SymTFT, Kaidi:NonInvTwist, Lee:MatchingHigher, Liu:2024znj, Morrison:5D_higher_form, Tian:2021cif, Tian:2024dgl, Tian:2025yrj,DelZotto:2015isa}. The interested readers can also consult the excellent reviews~\cite{Schafer-Nameki:ICTP, Bhardwaj:lecture, Luo:review, Gomes:review, Cordova:2022ruw, Shao:2023gho}. It was found that a particular effective tool for analyzing global symmetries of a QFT is the method of Symmetry Topological Field Theory (SymTFT)~\cite{Witten:1998wy, Freed:2012bs, Apruzzi:SymTFT, Kong:2020cie, Gaiotto:2020iye, vanBeest:SymTFT3D, Kaidi:SymTFT, Baume:2023kkf, Cvetic:2024dzu, Chen:2023qnv, Tian:2024dgl, Jia:2025jmn,Heckman:2024zdo,Braeger:2025rov,Heckman:2025lmw}. For a QFT $\mathcal{T}_\mathcal{F}$ with global symmetry $\mathcal{F}$ defined on spacetime $M_d$, the gist of SymTFT is to separate the symmetry and the dynamics of $\mathcal{T}_\mathcal{F}$ at two boundaries of a space homeomorphic to $M_d\times [0,1]$, where the dynamics of $\mathcal{T}_\mathcal{F}$ is encoded by the boundary conditions at the physical boundary on $M_d\times\{1\}\cong M_d$ while the symmetry is encoded by the boundary conditions at $B_{\textrm{top}} := M_d\times\{0\}$.

Although the von Neumann algebra of observables and the global symmetries of QFT have long been studied mostly parallelly, recent advances have revealed deep connections between them~\cite{Harlow:2018tng, Casini:2020rgj, Casini:2021zgr, Benedetti:2022zbb, Benedetti:2022ofj, Witten:2023qsv, Benedetti:2023owa, Benedetti:2024dku,Benedetti:2024utz, Yu:2025iqf}. Of particular relevance to the study in this paper is the recent work~\cite{Shao:2025mfj}, where the authors explored the connections between the algebraic approach and non-invertible symmetries in 2D. Their analysis focuses on the interplay between the existence of certain non-local operators and certain key locality properties of the symmetric sector von Neumann algebra $\mathcal{A}_{\mathcal{F}}$ (to be defined later) of diagonal RCFT and lattice examples. In this work, we aim to sharpen and generalize the result of~\cite{Shao:2025mfj} to arbitrary dimensions, in which case we find that the SymTFT perspective is particularly useful in proving the existence of the aforementioned specific non-local operators.

A careful reader may have already noticed an important subtlety. On one hand, for an ordinary QFT, the algebraic approach is carried out by assigning an algebra $\mathcal{A}(R)$ of observables in each spacetime region $R$. The way of this assignment is in principle part of the definition of a QFT and reflects the deep structures and properties of it. On the other hand, the generalized symmetry emphasizes the role of non-local, in particular extended higher-dimensional, operators. Therefore, an important question is whether such non-local operators living in $R$ should be included in the von Neumann algebra $\mathcal{A}(R)$. For this, we will follow the choice made in~\cite{Shao:2025mfj} where the non-local operators are included in $\mathcal{A}(R)$ as long as they do not extend to the boundary of spacetime. This constraint is imposed to exclude the defects that modify the whole Hilbert space rather than being an operator acting on the states in the Hilbert space. One should keep in mind that for a QFT defined alternatively, e.g. by a Hamiltonian or a Lagrangian, there is some freedom in the assignment of $\mathcal{A}(R)$ hence is entitled study the properties of various choices, regardless of whether there is a ``correct'' one~\cite{Shao:2025mfj}.

Another subtlety worth noting is that, unlike the full von Neumann algebra $\mathcal{A}$, $\mathcal{A}_{\mathcal{F}}$ by itself may not yield a well-defined theory. For example, the torus partition function of $\mathbb{Z}_2$-even sector of 2d Ising CFT is
    \begin{equation}
        Z_{\textrm{Ising}}^{\textrm{even}} = \textrm{Tr} P_+ e^{-\beta \hat{H}}\,,\quad P_+ = \frac{1}{2}\left(1+\hat{U}_{\mathbb{Z}_2} \right)\,,
    \end{equation}
with $P_+$ the projection operator and $\hat{U}_{\mathbb{Z}_2}$ the $\mathbb{Z}_2$ generator. It is obvious that the partition function is not invariant under the modular $S$-transformation, which brings in states in the $\mathbb{Z}_2$-twist sector and cannot be defined on general 2d manifolds. However, the symmetric sector of a theory is still an interesting object to study. It was pointed out in \cite{Haag:1996hvx} that two crucial properties of the algebra of observables, namely \emph{additivity} and \emph{Haag duality}, cannot hold simultaneously in the local $\mathbb{Z}_2$-even sector. This observation motivates the authors of~\cite{Shao:2025mfj} to further study how these two properties are preserved and violated in the symmetric sector von Neumann algebra $\mathcal{A}_{\mathcal{F}}$ of other 2d QFTs with the global symmetry $\mathcal{F}$. Motivated by the same reason, in this work we also focus on these properties of $\mathcal{A}_{\mathcal{F}}$.

The paper is organized as follows. We will first state the main results in section~\ref{sec:MainResults}. In section~\ref{sec:Symm_vNA}, we briefly review the basic notions of von Neumann algebra with an emphasis on its additivity and Haag duality, which are the main properties of von Neumann algebra of interest to us. After that, we define the symmetric sector von Neumann algebra $\mathcal{A}_\mathcal{F}$. In section~\ref{sec:SvNA_OpSpec}, we study the operator content of the symmetric sector von Neumann algebra of $\mathcal{T}_\mathcal{F}$ with an emphasis on the uncharged bi-local and the uncharged bi-twist operators (to be defined in that section), both of which will play crucial roles in later discussions. In section~\ref{sec:SvNA_SymTFT}, we use the SymTFT method to construct the bi-local and the bi-twist operators of $\mathcal{T}_\mathcal{F}$ and relate the existence of these operators to certain properties of the \emph{Lagrangian algebra} $\mathcal{L}$ of the corresponding SymTFT. In section~\ref{sec:Add_Haag_Violation}, we follow the construction of~\cite{Shao:2025mfj} to show that additivity and Haag duality of $\mathcal{A}_\mathcal{F}$ are violated in the presence of bi-local and bi-twist operators respectively, hence generalizing the results of~\cite{Shao:2025mfj} to arbitrary dimensions. Combining with the SymTFT construction in section~\ref{sec:SvNA_SymTFT}, we formulate a theorem relating the violation of additivity and Haag duality of $\mathcal{A}_\mathcal{F}$ to the properties of $\mathcal{L}$. Finally, in section~\ref{sec:examples}, we present several concrete examples of 2d QFT, including the diagonal RCFT, where we prove the equivalence of the results of \cite{Shao:2025mfj} and ours.

\subsection{Main results}\label{sec:MainResults}

In this work, we study a $d$-dimensional QFT $\mathcal{T}_{\mathcal{F}}$ with (non)-invertible 0-form symmetry $\mathcal{F}$ together with its $(d-2)$-form dual symmetry and focus on its symmetric sector von Neumann algebra $\mathcal{A}_{\mathcal{F}}(R)$ attached to a general spacetime region $R$. Namely, we focus on the local and non-local operators living in $R$ that are invariant under the action of any symmetry generator in $\mathcal{F}$.

Motivated by the 2d cases studied in~\cite{Shao:2025mfj}, where the violation of additivity or Haag duality of $\mathcal{A}_{\mathcal{F}}$ is determined by the properties of the symmetry algebra $\mathcal{F}$, we expect that the possible violation of additivity or Haag duality of $\mathcal{A}_{\mathcal{F}}$ for $\mathcal{T}_\mathcal{F}$ in higher dimension can similarly be determined using only the properties of $\mathcal{F}$. Therefore, it is natural to expect that they can be formulated in the language of SymTFT. The topological boundary condition at $B_{\textrm{top}}$ and the symmetry $\mathcal{F}$ are both determined by an algebraic object $\mathcal{L}$ in the SymTFT which, generalizing the \emph{Lagrangian algebra} for $d=2$, encodes all operators in the SymTFT that can simultaneously end on $B_{\textrm{top}}$. Thus we will call $\mathcal{L}$ the Lagrangian algebra of the SymTFT in general dimensions. For our purpose, we introduce the subalgebra $\widetilde{\mathcal{L}}$ which includes only the bulk extended operators in $\mathcal{L}$ that cannot connect to any non-trivial element of $\mathcal{F}$ supported on any submanifold of $B_{\textrm{top}}$~\footnote{We will define this in section~\ref{sec:SvNA_SymTFT}.}. In this work, we will focus only on the 1d and the $(d-1)$-d extended operators in the bulk which can possibly connect to 1d and $(d-1)$-d extended objects on $B_{\textrm{top}}$, respectively. We will also assume that the \emph{Totalitarian principle} holds, i.e. all allowed representations of $\mathcal{F}$ appear in the physical spectrum of $\mathcal{T}_\mathcal{F}$. Given the above assumptions, for the symmetric sector von Neumann algebra $\mathcal{A}_{\mathcal{F}}$ our main results are
\begin{itemize}
    \item The \emph{additivity} of $\mathcal{A}_{\mathcal{F}}$ is violated if $\widetilde{\mathcal{L}} \neq W_0 \oplus U_0$, where $W_0$ and $U_0$ are respectively the identity line operator and the identity $(d-1)$-dimensional extended operator in the SymTFT.
    \item The \emph{Haag duality} of $\mathcal{A}_{\mathcal{F}}$ is violated if $\widetilde{\mathcal{L}}\neq \mathcal{L}$.
\end{itemize}

When restricted to 2d diagonal RCFT, the above two criteria are respectively equivalent to the existence of invertible elements and non-invertible elements in $\mathcal{F}$, which in turn match the criteria in~\cite{Shao:2025mfj} under which additivity and Haag duality are violated in 2d diagonal RCFT, respectively. We find that, most generally the existence of an invertible element in $\mathcal{F}$ is not sufficient for additivity to be violated, for which we present a counterexample in 2d with $\textrm{Rep}(S_3)$ symmetry in section~\ref{sec:examples}.

\section{Additivity and Haag Duality of von Neumann Algebra}
\label{sec:Symm_vNA}

In this section, we review several basic facts about von Neumann algebras with an emphasis on their additivity and Haag duality. The readers can consult~\cite{Haag:1962, Haag:1963dh, Haag:1996hvx, Araki:1999ar}. We study QFT $\mathcal{T}_\mathcal{F}$ in spacetime $M_d$ whose global symmetries are characterized by 0-form symmetry algebra $\mathcal{F}$. Without loss of generality, we assume the spacetime is locally $M_{d-1}\times \mathbb{R}$ with Cauchy surface $M_{d-1}$.

The von Neumann algebra $\mathcal{A}$ of a QFT is defined to be a weakly closed $^*$-subalgebra of bounded operators $\mathcal{B}(\mathcal{H})$ on the Hilbert space $\mathcal{H}$ of the QFT which contains the unit operator~\cite{Haag:1996hvx}. The \emph{commutant} of a set of bounded operators $S$ is defined to be
\begin{equation}\label{eq:Commutant_Def}
    S' = \{a' \in \mathcal{B}(\mathcal{H}) | [a', a] = 0,\ \forall a \in S\}\,.
\end{equation}
A useful fact about a von Neumann algebra $\mathcal{A}$ is that it satisfies~(Theorem 2.1.4 of~\cite{Haag:1996hvx})
\begin{equation}
    \mathcal{A}'' = \mathcal{A}\,.
\end{equation}
A QFT is characterized by a net of assignments of the algebras
\begin{equation}
    R \rightarrow \mathcal{A}(R)
\end{equation}
where $R$ is an open, finitely-extended region of $M_d$. For a region $R$, we define $R'$ to be the \emph{causal complement} of $R$ which is the set of all points that are space-like separated from all points of $R$. The causality structure is expressed by the facts that $\mathcal{A}(R_1)$ commutes with $\mathcal{A}(R_2)$ when $R_1$ and $R_2$ are space-like separated, and $\mathcal{A}(R)\subset \mathcal{A}(R'')$ where $R''$ is the \emph{causal completion} of $R$.  We will restrict ourselves to the assignment $R\rightarrow\mathcal{A}(R)$ for $R$ a subset of a \emph{Cauchy surface} of the spacetime which is $M_{d-1}$ by our assumption.

The von Neumann algebra $\mathcal{A}(M_{d-1})$ has several properties due to the requirement of locality of QFT. Following~\cite{Shao:2025mfj}, in this work we shall focus on two of them, \emph{additivity} and \emph{Haag duality}~\cite{Haag:1963dh}. The von Neumann algebra is said to satisfy additivity if ((III.1.2) of~\cite{Haag:1996hvx})
\begin{equation}\label{eq:additivity}
	\mathcal{A}(R_1\cup R_2) = (\mathcal{A}(R_1)\cup \mathcal{A}(R_2))''
\end{equation}
for two disjoint regions $R_1$ and $R_2$ of $M_{d-1}$~\footnote{The additivity of $\mathcal{A}$ is can be written as $\mathcal{A}(R_1\vee R_2) = (\mathcal{A}(R_1)\cup \mathcal{A}(R_2))''$ ((III.4.8 of~\cite{Haag:1996hvx})) where $R_1\vee R_2$ is the causal completion of the union of two causally complete sets $R_1$ and $R_2$. When $R_1$ and $R_2$ and space-like separated we have $R_1\vee R_2 = R_1\cup R_2$~\cite{Casini:2002ah}. Therefore when $R_1$ and $R_2$ are two disjoint regions on a Cauchy surface we have $R_1\vee R_2 = R_1\cup R_2$, hence~(\ref{eq:additivity}) holds.}. On the other hand, the von Neumann algebra is said to satisfy Haag duality if ((III.4.9) of~\cite{Haag:1996hvx})
\begin{equation}\label{eq:Haag}
	\mathcal{A}(R') = \mathcal{A}(R)'
\end{equation}
for a region $R\in M_{d-1}$ and its complement $R'$~\footnote{Strictly speaking, this holds only for \emph{diamonds}, i.e. the causal completion of a ball in $M_{d-1}$~\cite{Haag:1996hvx}. We will see in section~\ref{sec:Add_Haag_Violation} that all the regions we consider in this work are indeed of this type.}. As we have said in the introduction, we will follow the choice of~\cite{Shadchin:2004yx} to include certain non-local operators in $\mathcal{A}(M_{d-1})$ and check the consequences they lead to for additivity and Haag duality. Roughly speaking, the additivity means that the von Neumann algebra in $R_1 \cup R_2$ is generated by algebras in $R_1$ and $R_2$, and the Haag duality means that the algebra of the causal complement is equal to the commutant of the algebra.

Of particular interest is what we shall call in this work the \emph{symmetric sector von Neumann algebra} defined as
\begin{equation}\label{eq:symm_vNA}
	\mathcal{A}_\mathcal{F}(M_{d-1}) = \{ \mathcal{O}\in\mathcal{A}(M_{d-1}) | N\mathcal{O} = \mathcal{O}N,\ \forall N\in \mathcal{F} \}\,,
\end{equation}
where we do not distinguish an element $N$ of $\mathcal{F}$ and its representation on the Hilbert space $\mathcal{H}$ of $\mathcal{T}_\mathcal{F}$. When the global symmetry on $\mathcal{H}$ is a representation of group $G$, $\mathcal{A}_G(M_{d-1})$ is nothing but the set of $G$-neutral operators and forms a closed subalgebra of $\mathcal{A}(M_{d-1})$. Thus $\mathcal{A}_\mathcal{F}(M_{d-1})$ can be viewed as a generalization of $\mathcal{A}_G(M_{d-1})$ to arbitrary fusion algebra $\mathcal{F}$ which can contain both invertible and non-invertible elements. We will see in section~\ref{sec:Add_Haag_Violation} that additivity and Haag duality of $\mathcal{A}_\mathcal{F}(M_{d-1})$ can be violated in the presence of certain non-local neutral operators in $\mathcal{A}_\mathcal{F}$, whose existence is completely determined by $\mathcal{F}$. Therefore, our task is to first investigate what types of neutral non-local operators are there in $\mathcal{A}_\mathcal{F}$.

\section{Operator Content of Symmetric Sector von Neumann Algebra}
\label{sec:SvNA_OpSpec}

In this section, we study various (non-local) operators in $\mathcal{A}_\mathcal{F}$ for the QFT $\mathcal{T}_\mathcal{F}$ defined by~(\ref{eq:symm_vNA}) using only the properties of the 0-form symmetry $\mathcal{F}$, which is most generally described by a fusion algebra with fusion rule
\begin{equation}\label{eq:fusion_rule}
    N_i \times N_j = \sum_k \mathcal{N}_{ij}^k N_k,\quad N_i\,,N_j\,,N_k\in \mathcal{F}\,,
\end{equation}
where $\mathcal{N}_{ij}^k \in \mathbb{Z}_{\geq0}$ are the fusion coefficients. An object $N$ is invertible if there exists an object $M\in\mathcal{F}$ such that $N\times M = 1$. As the local operators in $\mathcal{A}_{\mathcal{F}}$ are nothing but the uncharged local operators, there exist more non-trivial non-local operators in $\mathcal{A}_\mathcal{F}$.

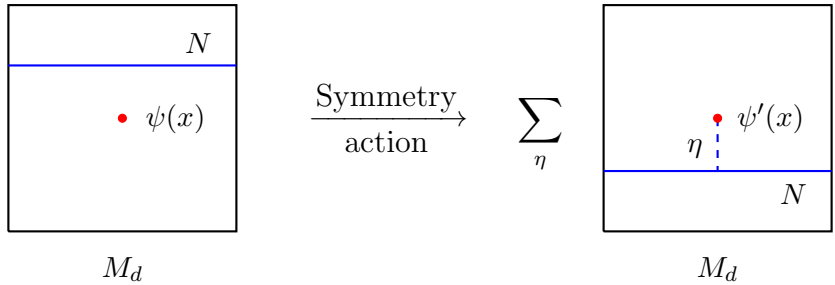
\begin{figure}[!h]
    \begin{equation}
            \begin{gathered}
        \begin{tikzpicture}
            \draw[thick] (0,0)--(0,3)--(3,3)--(3,0)--(0,0);
            \filldraw[red] (1.5,1.5) circle (1.5pt); 
            \draw[thick,blue] (0,2.2)--(3,2.2);
            \node at (2.2,1.5) {$\psi(x)$};
            \node at (2.5,2.5) {$N$};
            \node at (1.5,-0.5) {$M_{d}$};
            \node at (5,1.5) {$\xrightarrow[\textrm{\large action}]{\textrm{\large Symmetry}}$};
            \node at (7,1.3) {\large $\displaystyle \sum_{\eta}$};
        \end{tikzpicture}
    \end{gathered} \quad 
            \begin{gathered}
        \begin{tikzpicture}
            \draw[thick] (0,0)--(0,3)--(3,3)--(3,0)--(0,0);
            \draw[thick,blue] (0,0.8)--(3,0.8);
            \draw[thick,blue,dashed] (1.5,1.5)--(1.5,0.8);
            \filldraw[red] (1.5,1.5) circle (1.5pt); 
            \node at (2.2,1.5) {$\psi'(x)$};
            \node at (2.5,0.5) {$N$};
            \node at (1.2,1.1) {$\eta$};
            \node at (1.5,-0.5) {$M_{d}$};
        \end{tikzpicture}
    \end{gathered}\nonumber
    \end{equation}
    \caption{The action of symmetry operator $N\in\mathcal{F}$ on a charged operator. The symmetry can be invertible or non-invertible.}
    \label{fig:GeneralAction}
\end{figure}

Generally, for any local operator $\psi(x)\in \mathcal{A}$~\footnote{There is a subtlety that strictly speaking a local operator is unbounded, therefore technically one has to smear it out~\cite{Sorce:2023fdx}. Nevertheless this subtlety will not affect the discussions in this work, we therefore insist on calling it a local operator for simplicity.} suppported on a Cauchy surface $\Sigma$ which transforms non-trivially under $N\in\mathcal{F}$, we have
\begin{equation}\label{eq:general_symmetry_action}
    U(N)\psi(x) = \rho(N) \psi(x) U(N)\,.
\end{equation}
Without loss of generality, we assume that the local operator is located at $x^0 = 0$ and $N$ on the LHS of~(\ref{eq:general_symmetry_action}) is supported on a Cauchy surface at $x^0 > 0$ and that on the RHS is supported on a Cauchy surface at $x^0 < 0$ as illustrated in Figure~\ref{fig:GeneralAction}. In general, the action $\rho(N)$ can be very complicated and does not have to be a linear transformation.
Actually, for a non-invertible symmetry, various defects $\eta$ can be generated when $N$ passes across $\psi(x)$ and all possible configurations need to be summed over. In this sense, $\rho(N)$ in~(\ref{eq:general_symmetry_action}) is a condensed notation for this complicated action. On the other hand, no defects will be generated when an invertible symmetry generator passes across $\psi(x)$, in which case $\rho(N)$ acts as a linear transformation on $\psi(x)$.

We first consider an invertible element $N\in\mathcal{F}$. Actually, all invertible elements of a general ring $\mathcal{F}$ must from a group $G\subset \mathcal{F}$, and one can assume that there exists local field $\psi^i_{\mathbf{r}}(x)$ living in the vector space $V_\mathbf{r}$ of a representation $\mathbf{r}$ of $G$. Concretely, the symmetry action~(\ref{eq:general_symmetry_action}) of $N$ can be characterized by
\begin{equation}
    \rho(N) \in GL(V_\mathbf{r}) : V_\mathbf{r}\rightarrow V_\mathbf{r} \,.
\end{equation}
Therefore, in the superselection sector $V_\mathbf{r}$,~(\ref{eq:general_symmetry_action}) can be written as
\begin{equation}\label{eq:inv_N_on_Vr}
    N \psi_\mathbf{r}^i(x) = \rho_\mathbf{r}(N)^i_{\ j} \psi_\mathbf{r}^j(x)N\,.\quad (i,j=1,\cdots,\dim V_\mathbf{r})\
\end{equation}
Using~(\ref{eq:inv_N_on_Vr}) and the fact that in an ordinary QFT the existence of a local operator that transforms non-trivially under the action of $N$ also guarantees the existence of its charge conjugation living in the dual vector space $\overline{V_\mathbf{r}}$, one can construct the following bi-local operator
\begin{equation}\label{eq:bi-local_ops}
    \tr \psi_\mathbf{r}(x)\psi_{\overline{\mathbf{r}}}(y) := \psi_\mathbf{r}^i(x) \psi_{\overline{\mathbf{r}}i}(y)\,,
\end{equation}
for $\psi_\mathbf{r}^i(x) \in V_\mathbf{r}$ and $\psi_{\overline{\mathbf{r}}i}(x) \in \overline{V_\mathbf{r}}$. It is not hard to check that
\begin{equation}\label{eq:Inv_bi-local}
    N ( \tr\psi_\mathbf{r}(x)\psi_{\overline{\mathbf{r}}}(y) ) = (\psi_\mathbf{r}^i(x)  \rho_\mathbf{r}(N)^j_{\ i} \rho_{\overline{\mathbf{r}}}(N)_j^{\ k} \psi_{\overline{\mathbf{r}}k}(y)) N = ( \tr\psi_\mathbf{r}(x)\psi_{\overline{\mathbf{r}}}(y) ) N\,.
\end{equation}
Therefore, the bi-local operator $\tr\psi_\mathbf{r}(x)\psi_{\overline{\mathbf{r}}}(y)$ is invariant under the action of $N\in G$. When $\mathcal{F}$ is a group, the local operator $\psi^i_{\mathbf{r}}(x)$ lives in a representation of $\mathcal{F}$ and the bi-local operator~(\ref{eq:bi-local_ops}) is invariant under all action of elements of $\mathcal{F}$, hence is in $\mathcal{A}_\mathcal{F}(M_{d-1})$.

\begin{figure}[!h]
    \begin{equation}
            \begin{gathered}
        \begin{tikzpicture}
            \draw[thick] (0,0)--(0,3)--(3,3)--(3,0)--(0,0);
            \filldraw[red] (0.8,1.5) circle (1.5pt); 
            \filldraw[red] (2.2,1.5) circle (1.5pt); 
            \draw[thick,blue] (0,2.2)--(3,2.2);
            \node at (0.9,1) {$\psi(x)$};
            \node at (2.3,1) {$\overline{\psi}(y)$};
            \node at (2.5,2.5) {$N$};
            \node at (1.5,-0.5) {$M_{d}$};
            \node at (4,1.5) {$\xrightarrow[\textrm{action}]{\textrm{Symmetry}}$};
        \end{tikzpicture}
    \end{gathered} \  
            \begin{gathered}
        \begin{tikzpicture}
            \draw[thick] (0,0)--(0,3)--(3,3)--(3,0)--(0,0);
            \draw[thick,blue] (0,0.8)--(3,0.8);
            \draw[thick,blue,dashed] (0.8,1.5)--(0.8,0.8);
            \draw[thick,blue,dashed] (2.2,1.5)--(2.2,0.8);
            \filldraw[red] (0.8,1.5) circle (1.5pt); 
            \filldraw[red] (2.2,1.5) circle (1.5pt); 
            \node at (0.9,2) {$\psi(x)$};
            \node at (2.3,2) {$\overline{\psi}(y)$};
            \node at (2.5,0.5) {$N$};
            \node at (0.6,1.1) {$\eta$};
            \node at (2.4,1.1) {$\eta$};
            \node at (1.5,-0.5) {$M_{d}$};
            \draw[thick,dashed,blue,->] (0.8,0.6)--(1.35,0.6);
            \draw[thick,dashed,blue,->] (2.2,0.6)--(1.65,0.6);
            \node at (4.2,1.5) {$\xrightarrow{\textrm{Deformation}}$};
        \end{tikzpicture}
    \end{gathered}\ 
            \begin{gathered}
        \begin{tikzpicture}
            \draw[thick] (0,0)--(0,3)--(3,3)--(3,0)--(0,0);
            \draw[thick,blue] (0,0.8)--(3,0.8);
            \draw[thick,blue,dashed] (0.8,1.5)--(2.2,1.5);
            \filldraw[red] (0.8,1.5) circle (1.5pt); 
            \filldraw[red] (2.2,1.5) circle (1.5pt); 
            \node at (0.9,2) {$\psi(x)$};
            \node at (2.3,2) {$\overline{\psi}(y)$};
            \node at (1.5,1.2) {$\eta$};
            \node at (2.5,0.5) {$N$};
            \node at (1.5,-0.5) {$M_{d}$};
        \end{tikzpicture}
    \end{gathered}   \nonumber
    \end{equation}
    \caption{Symmetry action of a non-invertible element on a bi-local operator where a defect line $D_{\ell}$ connecting $x$ and $y$ is generated upon deformation.}
    \label{fig:Noninv_bi-local}
\end{figure}
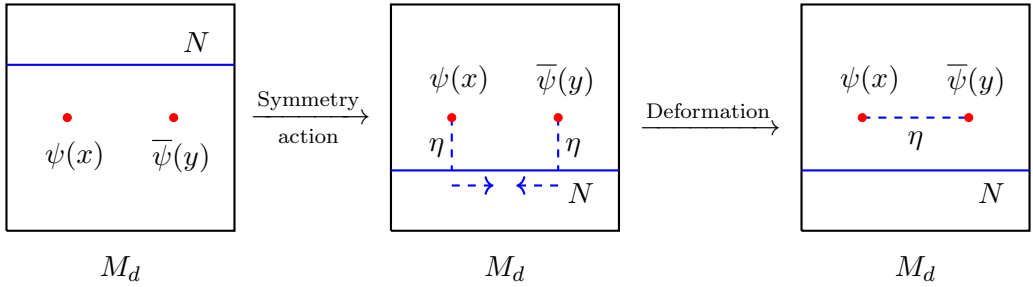

The situation becomes tricky when there exist non-invertible elements in $\mathcal{F}$, since a representation of $G\subset\mathcal{F}$ does not fully characterize the transformation of $\psi(x)$ under $\mathcal{F}$. As illustrated in Figure~\ref{fig:GeneralAction}, typically under the action of a non-invertible symmetry generator $N$, a defect line $\eta$ is generated, which stretches between the local operator $\psi(x)$ and the support of the symmetry operator. Therefore, the bi-local operator $\tr \psi(x)\overline{\psi}(y)$ built upon such $\psi(x)$ cannot be invariant under $N$, rather a defect line $\eta(\gamma_{xy})$ connecting $\psi(x)$ and $\overline{\psi}(y)$ is generated in the process illustrated in Figure~\ref{fig:Noninv_bi-local}. Hence, such a bi-local operator is not an element of $\mathcal{A}_\mathcal{F}(M_{d-1})$.
A prominent example is the order operator $\sigma$ with conformal wright $(h,\overline{h}) = (1/16,1/16)$ in Ising CFT~\cite{Oshikawa:1996dj, Petkova:2000ip, Frohlich:2004ef}, which is odd under the invertible $\mathbb{Z}_2$-symmetry generator, but transforms non-trivially in the fashion illustrated in Figure~\ref{fig:GeneralAction} under the non-invertible object of the symmetry category of Ising CFT. Therefore, it becomes rather tricky to determine if a bi-local operator is in $\mathcal{A}_{\mathcal{F}}(M_{d-1})$ when $\mathcal{F}$ is non-invertible. We will see in the next section that this can be determined by a simple criterion via the SymTFT method.

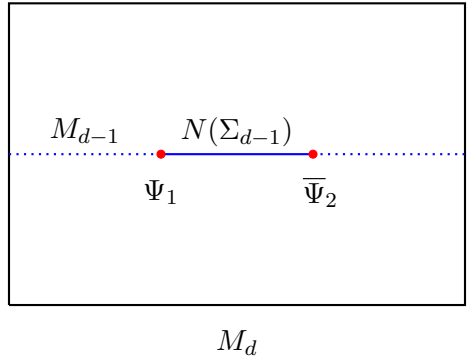
\begin{figure}[!h]
    \begin{equation}
            \begin{gathered}
        \begin{tikzpicture}
            \draw[thick] (0,0)--(0,4)--(6,4)--(6,0)--(0,0);
            \draw[thick,blue] (2,2)--(4,2);
            \draw[thick,dotted,blue] (0,2)--(2,2);
            \draw[thick,dotted,blue] (4,2)--(6,2);
            \filldraw[red] (2,2) circle (1.5pt); 
            \filldraw[red] (4,2) circle (1.5pt); 
            \node at (2,1.5) {$\Psi_1$};
            \node at (4.1,1.5) {$\overline{\Psi}_2$};
            \node at (3,2.3) {$N(\Sigma_{d-1})$};
            \node at (1,2.3) {$M_{d-1}$};
            \node at (3,-0.5) {$M_{d}$};
        \end{tikzpicture}
    \end{gathered} \nonumber
    \end{equation}
    \caption{A twist operator supported on $\Sigma_{d-1}\subset M_{d-1}$ with dressing $\Psi_1(\Sigma_{d-2}) $ and $ \Psi_2(\Sigma'_{d-2})$ at the boundary of $\Sigma_{d-1}$.}
    \label{fig:TwistOperator}
\end{figure}

Allowing non-local operators in $\mathcal{A}_{\mathcal{F}}(M_{d-1})$ also opens up the possibility of having more general extended operators beyond bi-local operators. In general, there exist twist operators of the form $\Psi(\partial\Sigma_{d-1})N(\Sigma_{d-1})$ where $N(\Sigma_{d-1})$ is a symmetry generator $N\in\mathcal{F}$ supported on $\Sigma_{d-1}\in M_{d-1}$ and $\Psi(\partial\Sigma_{d-1})$ is a dressing on the boundary $\partial \Sigma_{d-1}$. A prominent example is the operator $\psi(x)L_\epsilon\psi(y)$ in Ising CFT, where $\psi$ is a left-moving fermion with conformal weight $(h,\overline{h}) = (1/2,0)$ and $L_\epsilon$ the invertible $\mathbb{Z}_2$-symmetry generator supported on a segment connecting $x$ and $y$ in the 1D space $S^1$~\footnote{It is argued in~\cite{Shao:2025mfj} that this operator is in $\mathcal{A}_\mathcal{F}(S^1)$ for $\mathcal{F}$ the symmetry category of Ising CFT.}. Generalizing such 2d twist operator, we assume that $M_{d-1}$ is locally $M_{d-2}\times \mathbb{R}$ and consider an element $N\in \mathcal{F}$ supported on $\Sigma_{d-1} \cong M_{d-2}\times (0,1)$ denoted by $N(\Sigma_{d-1})$. We then dress $N(\Sigma_{d-1})$ by $\Psi_1$ supported on $\Sigma_{d-2}=M_{d-2}\times \{0\}$ and $\Psi_2$ supported on $\Sigma'_{d-2}=M_{d-2}\times \{1\}$ to form what we call a \emph{bi-twist} operator $\Psi_1(\Sigma_{d-2}) N(\Sigma_{d-1}) \Psi_2(\Sigma'_{d-2})$, which is plotted schematically in Figure~\ref{fig:TwistOperator}. To avoid clustering symbols, we will omit the supports in $\Psi_1(\Sigma_{d-2})$ and $\Psi_2(\Sigma'_{d-2})$ when it does not cause any confusion.

\begin{figure}[!h]
    \begin{equation}
            \begin{gathered}
        \begin{tikzpicture}
            \draw[thick] (0,0)--(0,4)--(6,4)--(6,0)--(0,0);
            \draw[thick,blue] (3,2)--(3,0);
            \draw[thick,dotted,blue] (0,3)--(6,3);
            \draw[thick,dotted,blue] (0,2)--(6,2);
            \draw[thick,dotted,blue] (0,1)--(6,1);
            \filldraw[red] (3,2) circle (1.5pt); 
            \node at (4,1.7) {$\Psi(\partial \Sigma_{d-1})$};
            \node at (3.85,0.45) {$N(\Sigma_{d-1})$};
            \node at (2.25,0.5) {$x^1=0$};
            \node at (0.8,2.3) {$x^0=0_{\ \,}$};
            \node at (0.8,3.3) {$x^0=0_+$};
            \node at (0.8,1.3) {$x^0=0_-$};
            \node at (3,-0.5) {$M_{d}$};
        \end{tikzpicture}
    \end{gathered} \nonumber
    \end{equation}
    \caption{A twist operator supported on $\Sigma$ with dressing $\Psi$ at $\partial\Sigma$ to ensure consistent termination of $N$.}
    \label{fig:BC}
\end{figure}
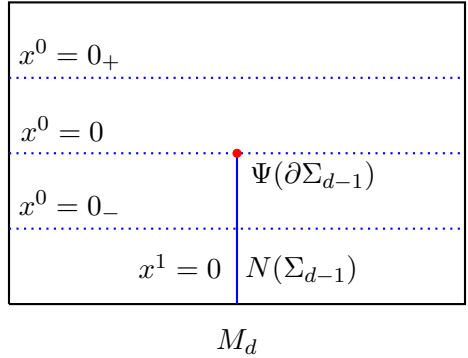

Before checking if there exists any bi-twist operator $\Psi_1 N(\Sigma_{d-1}) \Psi_2$ in $\mathcal{A}_\mathcal{F}(M_{d-1})$, we need to check its existence in $\mathcal{A}(M_{d-1})$ first. We will make the discussion general by considering the twist operator $\Psi(\partial\Sigma_{d-1})N(\Sigma_{d-1})$ where $\Sigma_{d-1}$ is a general simply-connected open subset of $M_{d-1}$. Physically, $\Psi(\partial\Sigma_{d-1})$ is dressed on $N(\Sigma_{d-1})$ to ensure consistent termination of $N(\Sigma_{d-1})$ at $\partial\Sigma_{d-1}$. To see how this consistency condition constrains $\Psi$ for given $N$, we consider the configuration in Figure~\ref{fig:BC}.
For any local field $\phi(x)$ that is charged under $N$, we have
\begin{equation}\label{eq:N_termination}
    \begin{split}
        \phi(x^0 = 0_-, x^1 = 0_+, \vec{x}) &= N\cdot \phi(x^0 = 0_-, x^1 = 0_-, \vec{x}), \\
        \phi(x^0 = 0_+, x^1 = 0_+, \vec{x}) &= \phi(x^0 = 0_+, x^1 = 0_-, \vec{x})
    \end{split}
\end{equation}
where $\vec{x} = (x^2,\cdots,x^{d-1})$ and the precise representation of $N$ is not specified in order to keep complete generality. If we apply a global symmetry transformation $g$ such that $\phi \rightarrow \phi' =g\cdot\phi$, the boundary conditions will change to
\begin{equation}
    \begin{split}
        \phi'(x^0 = 0_-, x^1 = 0_+, \vec{x}) &= N'\cdot \phi'(x^0 = 0_-, x^1 = 0_-, \vec{x}), \\
        \phi'(x^0 = 0_+, x^1 = 0_+, \vec{x}) &= \phi'(x^0 = 0_+, x^1 = 0_-, \vec{x})\,,
    \end{split}
\end{equation}
with $N'=gNg^{-1}$, which implies the tail of the twist operator $\Psi(\partial\Sigma_{d-1})N(\Sigma_{d-1})$ will be transformed to $N'(\Sigma_{d-1})$. If we consider $N'=N$ so that $g$ is an element of the centralizer $C_\mathcal{F}(N)$ of $N$ in $\mathcal{F}$, then the defect $N$ remains the same and $\Psi$ is allowed to transform under a representation of (or more precisely, a module over) $C_\mathcal{F}(N)$. One may further use a module over $C_\mathcal{F}(N)$ to induce a module over $\mathcal{F}$.
% In other words, the invariant subspace of $N$ naturally yields a representation of (or more precisely, a module over) $C_\mathcal{F}(N)$. 
% To ensure agreement of the two conditions in~(\ref{eq:N_termination}) at $(x^0, x^1) = (0, 0)$, we require
% \begin{equation}\label{eq:consistent_termination}
%     \phi(x^0 = 0, x^1 = 0, \vec{x}) = N\cdot \phi(x^0 = 0, x^1 = 0, \vec{x})\,.
% \end{equation}
% Though $\phi(0,0,\vec{x})$ is invariant under $N$, it is still not completely fixed. For this we note that~(\ref{eq:consistent_termination}) allows $\phi(0,0,\vec{x})$ to transform under a representation of the centralizer $C_\mathcal{F}(N)$ of $N$ in $\mathcal{F}$. More precisely, for any $M\in\mathcal{F}$ such that $M\times N = N\times M$, we have
% \begin{equation}
%     M\cdot \phi(x^0 = 0, x^1 = 0, \vec{x}) = M\cdot N\cdot \phi(x^0 = 0, x^1 = 0, \vec{x}) = N\cdot \left( M\cdot \phi(x^0 = 0, x^1 = 0, \vec{x})\right)\,,
% \end{equation}
% hence $M\cdot \phi(x^0 = 0, x^1 = 0, \vec{x})$ is also invariant under $N$. In other words, the invariant subspace of $N$ naturally yields a representation of (or more precisely, a module over) $C_\mathcal{F}(N)$. On the other hand, one may use a module over $C_\mathcal{F}(N)$ to induce a module over $\mathcal{F}$. 
% Therefore, the boundary condition~(\ref{eq:consistent_termination}) at $\partial\Sigma_{d-1}$ is fully determined by a representation of $C_\mathcal{F}(N)$. 
Hence, for a twist operator $\Psi(\partial\Sigma_{d-1}) N(\Sigma_{d-1})$ to exist, we require $\Psi$ live in a module over $C_\mathcal{F}(N)$~\footnote{The cautious readers may note that this resembles the data of the \emph{Drinfeld center} of the category of $G$-graded vector spaces when $\mathcal{F} \cong G$~\cite{etingof2017tensor}. This is by no means a coincidence and will be studied in~\cite{Jia:ToAppear}.}. For a global symmetry described by a group $G$, this is equivalent to requiring $\Psi$ be in a vector space carrying a representation of $C_G(N)$ for $N\in G$.

We now investigate the conditions for $\Psi(\partial\Sigma_{d-1})N(\Sigma_{d-1})$ to be in $\mathcal{A}_\mathcal{F}(M_{d-1})$. Now we will restrict the discussion to the bi-twist operators $\Psi_1 N(\Sigma_{d-1}) \Psi_2$. For simplicity, we start with $N\in C(\mathcal{F})$, the set of elements of $\mathcal{F}$ that commute with everything else in $\mathcal{F}$. To construct a twist operator $\Psi_1N(\Sigma_{d-1})\Psi_2$ in $\mathcal{A}_{\mathcal{F}}(M_{d-1})$, we need to dress suitable $\Psi_i$'s at the boundary of $\Sigma_{d-1}$. As discussed earlier, each $\Psi_i$ is labeled by a module over $C_\mathcal{F}(N)$. Since we restrict to $N\in C(\mathcal{F})$, we have $C_{\mathcal{F}}(N) \cong \mathcal{F}$ hence each $\Psi_i$ is characterized by a module over $\mathcal{F}$. When $\mathcal{F}$ is invertible, i.e. a group, each $\Psi_i$ transforms linearly under all generators of $\mathcal{F}$, and we could simply choose $\Psi_1$ to be in representation $\mathbf{r}$ and $\Psi_2$ to be in the dual representation $\overline{\mathbf{r}}$. Contracting all the indices by taking the trace like in~(\ref{eq:bi-local_ops}) yields $\Psi_1N(\Sigma_{d-1})\Psi_2 \in \mathcal{A}_\mathcal{F}(M_{d-1})$. However, the above construction of uncharged bi-twist operators works only for invertible $\mathcal{F}$ with non-trivial center, and does not generalize to non-invertible $\mathcal{F}$ or $\mathcal{F}$ with trivial center. The point, though, is to show that there does exist $\mathcal{F}$ such that $\mathcal{A}_\mathcal{F}(M_{d-1})$ contains bi-twist operators. Again, we will see in the next section that uncharged bi-twist operators can be constructed conveniently via the SymTFT method.

\section{Symmetric Sector von Neumann Algebra from SymTFT}
\label{sec:SvNA_SymTFT}

\begin{figure}[h]
    \begin{equation}
            \begin{gathered}
        \begin{tikzpicture}
            \draw[thick] (0,0)--(0,3);
            \draw[thick] (0,3)--(2,3.5);
            \draw[thick] (2,3.5)--(2,0.5);
            \draw[thick] (2,0.5)--(0,0);
            % \draw[thick,red] (1,1.75)--(1,1.75+1.5);
            \filldraw[red] (1,1.75) circle (1.5pt); 
            % \draw[thick,blue] (1,1.75) ellipse (0.5 and 1);
            % \node at (1.5,3) {$\hat{W}$};
            % \node at (1.5,1.75+1) {$U(g)$};
            \node at (1.35,1.5) {$\psi$};
            \draw[thick,blue] (0,2)--(2,2.5);
            \node at (1.5,2.75) {$N$};
            \node at (1,-0.5) {$M_d$};
            \node at (4,1.75) {\Large $\Leftrightarrow$};
        \end{tikzpicture}\\
        {\textrm{\Large $\Updownarrow$}}\qquad \qquad \qquad \\
        \begin{tikzpicture}
            \draw[thick] (0,0)--(0,3);
            \draw[thick] (0,3)--(2,3.5);
            \draw[thick] (2,3.5)--(2,0.5);
            \draw[thick] (2,0.5)--(0,0);
            % \draw[thick,red] (1,1.75)--(1,1.75+1.5);
            \filldraw[red] (1,1.75) circle (1.5pt); 
            % \draw[thick,blue] (1,1.75) ellipse (0.5 and 1);
            % \node at (1.5,3) {$\hat{W}$};
            % \node at (1.5,1.75+1) {$U(g)$};
            \node at (1.35,1.5) {$\psi$};
            \draw[blue,thick,dashed] (1,1.7)--(1,1);
            \draw[thick,blue] (0,0.75)--(2,1.25);
            \node at (1.5,0.75) {$N$};
            \node at (1,-0.5) {$M_2$};
            \node at (4,1.75) {\Large $\Leftrightarrow$};
        \end{tikzpicture}
    \end{gathered} \qquad \qquad
            \begin{gathered}
        \begin{tikzpicture}
            \draw[thick] (0,0)--(0,3);
            \draw[thick] (0,3)--(2,3.5);
            \draw[thick] (2,3.5)--(2,0.5);
            \draw[thick] (2,0.5)--(0,0);
            \draw[thick] (4,0)--(4,3);
            \draw[thick] (4,3)--(6,3.5);
            \draw[thick] (6,3.5)--(6,0.5);
            \draw[thick] (6,0.5)--(4,0);
            \draw[thick] (0,0)--(4,0);
            \draw[thick] (0,3)--(4,3);
            \draw[thick] (2,3.5)--(6,3.5);
            \draw[thick] (2,0.5)--(4,0.5);
            \draw[thick,dashed] (4,0.5)--(6,0.5);
            \draw[thick,red] (1,1.75)--(4,1.75);
            \draw[thick,red,dashed] (4,1.75)--(5,1.75);
            \filldraw[red] (1,1.75) circle (1.5pt);
            \filldraw[red] (5,1.75) circle (1.5pt); 
            % \draw[thick,blue] (1,1.75) ellipse (0.5 and 1);
            \node at (3,2) {$W$};
            \draw[thick,blue] (0,2)--(2,2.5);
            \node at (1.5,2.75) {$N$};
            \node at (5,1.25) {$\widetilde{\psi}$};
            \node at (1,-0.5) {$B_{\textrm{top}}$};
            \node at (5,-0.5) {$B_{\textrm{phys}}$};
            % \node at (1.5,3) {$\hat{W}$};
            % \node at (1.5,1.75+1) {$U(g)$};
        \end{tikzpicture}\\
        {\textrm{\Large $\Updownarrow$}}\\
        \begin{tikzpicture}
            \draw[thick] (0,0)--(0,3);
            \draw[thick] (0,3)--(2,3.5);
            \draw[thick] (2,3.5)--(2,0.5);
            \draw[thick] (2,0.5)--(0,0);
            \draw[thick] (4,0)--(4,3);
            \draw[thick] (4,3)--(6,3.5);
            \draw[thick] (6,3.5)--(6,0.5);
            \draw[thick] (6,0.5)--(4,0);
            \draw[thick] (0,0)--(4,0);
            \draw[thick] (0,3)--(4,3);
            \draw[thick] (2,3.5)--(6,3.5);
            \draw[thick] (2,0.5)--(4,0.5);
            \draw[thick,dashed] (4,0.5)--(6,0.5);
            \draw[thick,red] (1,1.75)--(4,1.75);
            \draw[thick,red,dashed] (4,1.75)--(5,1.75);
            \filldraw[red] (1,1.75) circle (1.5pt);
            \filldraw[red] (5,1.75) circle (1.5pt); 
            % \draw[thick,blue] (1,1.75) ellipse (0.5 and 1);
            \node at (3,2) {$W$};
            \draw[thick,blue] (0,0.75)--(2,1.25);
            \draw[blue,thick,dashed] (1,1.7)--(1,1);
            \node at (1.5,0.75) {$N$};
            \node at (5,1.25) {$\widetilde{\psi}$};
            \node at (1,-0.5) {$B_{\textrm{top}}$};
            \node at (5,-0.5) {$B_{\textrm{phys}}$};
            % \node at (1.5,3) {$\hat{W}$};
            % \node at (1.5,1.75+1) {$U(g)$};
        \end{tikzpicture}
    \end{gathered}\nonumber 
    \end{equation}
    \caption{The SymTFT description of a symmetry action of $\mathcal{T}_\mathcal{F}$.}
    \label{Fig-SymTFT}
\end{figure}
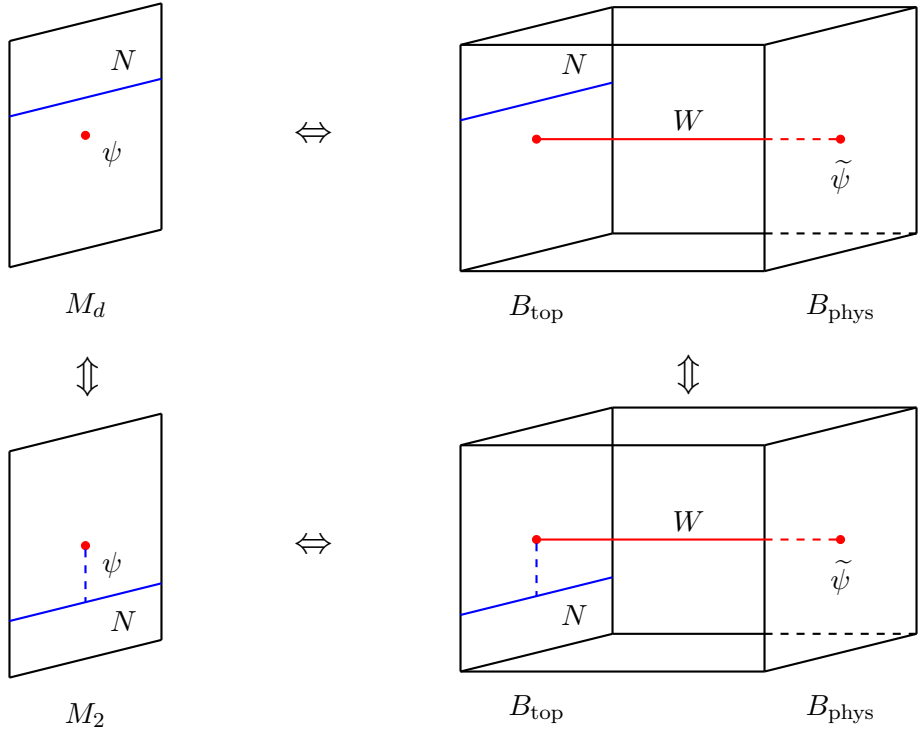

The bi-local and the bi-twist operators in the symmetric sector von Neumann algebra discussed in the last section can be studied in a more transparent and unified manner from the perspective of SymTFT. The idea of SymTFT is to attach $\mathcal{T}_\mathcal{F}$ with global $\mathcal{F}$-symmetry on $M_d$ to a topological field theory on $M_d\times [0,1]$ with two boundaries $B_{\textrm{top}}\cong M_d\times \{0\}$ and $B_{\textrm{phys}}\cong M_{d}\times \{1\}$, as depicted in Figure~\ref{Fig-SymTFT}. The data of the symmetry and the dynamics of $\mathcal{T}_{\mathcal{F}}$ are separately stored at the topological boundary $B_{\textrm{top}}$ and the physical boundary $B_{\textrm{phys}}$. A symmetry generator $N\in\mathcal{F}$ is supported on a $(d-1)$-dimensional submanifold of $B_{\textrm{top}}$ bordant to a Cauchy surface $M_{d-1}\subset B_{\textrm{phys}}$, and a local operator $\psi$ on $B_{\textrm{phys}}$ is extended into the bulk to be a line operator $W$ stretching between the two boundaries. A symmetry action of $N$ on a local operator $\psi$ in $\mathcal{T}_\mathcal{F}$ can be understood from the SymTFT perspective as $N$ on $B_\text{top}$ passing through the end point of the corresponding $W$ on $B_\text{top}$. We anticipate the possibility that a defect connecting $N$ and $W$ on $B_\text{top}$ can be generated (in particular when $\mathcal{F}$ is non-invertible), which is represented by the dashed line in Figure~\ref{Fig-SymTFT} (cf.~Figure~\ref{fig:GeneralAction}). 

To be concrete, we will first study 2d theory with symmetry $\mathcal{F}$, whose corresponding 3d SymTFT was first introduced in~\cite{TURAEV1992865,Barrett:1993ab}. The set of line operators (anyons) was studied in detail and shown to be in one-to-one correspondence with irreducible representations of $\mathcal{F}$~\cite{Izumi:2000qa,Evans:2010yr,MUGER2003159,Lin:2022dhv}. This fact plays an important role in the following discussion. We will review various types of (extended) operators in the SymTFT and discuss the conditions for them to be in the symmetric sector von Neumann algebra $\mathcal{A}_{\mathcal{F}}(M_{1})$. We will then generalize the discussion to $d > 2$.

\subsection{Uncharged Operators from SymTFT}

\subsubsection*{Local operators}

We begin with the local operators of $\mathcal{T}_\mathcal{F}$. A generic line operator in its corresponding SymTFT can be written as
\begin{equation}\label{eq:termitable_line}
    L_\alpha = n_{\alpha} W_{\alpha}\,,\quad (n_\alpha \in \mathbb{Z}_{\geq 0})
\end{equation}
where $W_\alpha$ is a line operator in the bulk SymTFT as shown in Figure~\ref{Fig-SymTFT}. The coefficient $n_\alpha$ will be explained later. A line operator $L_\alpha$ is in a \emph{Lagrangian algebra} $\mathcal{L}$, i.e. $L_\alpha\in\mathcal{L}$ (and $W_\alpha\in\mathcal{L}$), if it is allowed to end on $B_{\textrm{top}}$~\cite{Cong:2016ayp}, which in turn means the corresponding $W_\alpha$ is terminable at a point on $B_{\textrm{top}}$~\footnote{A line $W_\alpha$ is said to be not terminable on $B_{\textrm{top}}$ when it has to connect to a defect line, which will later be called a ``tail'', on $B_{\textrm{top}}$ extending to infinity. We will encounter this case in a moment.}. More precisely, one can write $\mathcal{L}$ as
\begin{equation}\label{eq:Lagrangian_algebra}
    \mathcal{L} = \bigoplus_\alpha L_\alpha\,,
\end{equation}
where $L_\alpha$'s mutually commute with each other and satisfy certain technical conditions~\cite{Cong:2016ayp, Bhardwaj:2023idu}. In particular, the identity line operator $W_0$ in the bulk is always in the direct sum~(\ref{eq:Lagrangian_algebra}) with $n_0=1$. Inequivalent topological boundary conditions at $B_{\textrm{top}}$ are in one-to-one correspondence with inequivalent choices of the Lagrangian algebras of the SymTFT. Moreover, $L_\alpha$ corresponds to a local operator of $\mathcal{T}_\mathcal{F}$ upon interval compactification.

\begin{figure}[!h]
    \begin{equation}
            \begin{gathered}
        \begin{tikzpicture}
            \draw[thick] (0,0)--(0,3);
            \draw[thick] (0,3)--(2,3.5);
            \draw[thick] (2,3.5)--(2,0.5);
            \draw[thick] (2,0.5)--(0,0);
            % \draw[thick,red] (1,1.75)--(1,1.75+1.5);
            \filldraw[red] (1,1.75) circle (1.5pt); 
            % \draw[thick,blue] (1,1.75) ellipse (0.5 and 1);
            % \node at (1.5,3) {$\hat{W}$};
            % \node at (1.5,1.75+1) {$U(g)$};
            \node at (1.35,1.5) {$\psi$};
            \draw[thick,blue] (0,2)--(2,2.5);
            \node at (1.5,2.75) {$N$};
            \node at (1,-0.5) {$M_2$};
            \node at (4,1.75) {\Large $\Leftrightarrow$};
        \end{tikzpicture}
    \end{gathered} \qquad \qquad
            \begin{gathered}
        \begin{tikzpicture}
            \draw[thick] (0,0)--(0,3);
            \draw[thick] (0,3)--(2,3.5);
            \draw[thick] (2,3.5)--(2,0.5);
            \draw[thick] (2,0.5)--(0,0);
            \draw[thick] (4,0)--(4,3);
            \draw[thick] (4,3)--(6,3.5);
            \draw[thick] (6,3.5)--(6,0.5);
            \draw[thick] (6,0.5)--(4,0);
            \draw[thick] (0,0)--(4,0);
            \draw[thick] (0,3)--(4,3);
            \draw[thick] (2,3.5)--(6,3.5);
            \draw[thick] (2,0.5)--(4,0.5);
            \draw[thick,dashed] (4,0.5)--(6,0.5);
            \draw[thick,red,dotted] (1,1.75)--(5,1.75);
            % \draw[thick,red,dashed] (4,1.75)--(5,1.75);
            \filldraw[red] (1,1.75) circle (1.5pt);
            \filldraw[red] (5,1.75) circle (1.5pt); 
            % \draw[thick,blue] (1,1.75) ellipse (0.5 and 1);
            \node at (3,2) {$W_0$};
            \draw[thick,blue] (0,2)--(2,2.5);
            \node at (1.5,2.75) {$N$};
            \node at (5.35,1.5) {$\psi$};
            \node at (1,-0.5) {$B_{\textrm{top}}$};
            \node at (5,-0.5) {$B_{\textrm{phys}}$};
            % \node at (1.5,3) {$\hat{W}$};
            % \node at (1.5,1.75+1) {$U(g)$};
        \end{tikzpicture}
    \end{gathered}\nonumber 
    \end{equation}
    \caption{An uncharged local operator seen from the SymTFT perspective.}
    \label{Fig-Symmetric-Local-Operator}
\end{figure}
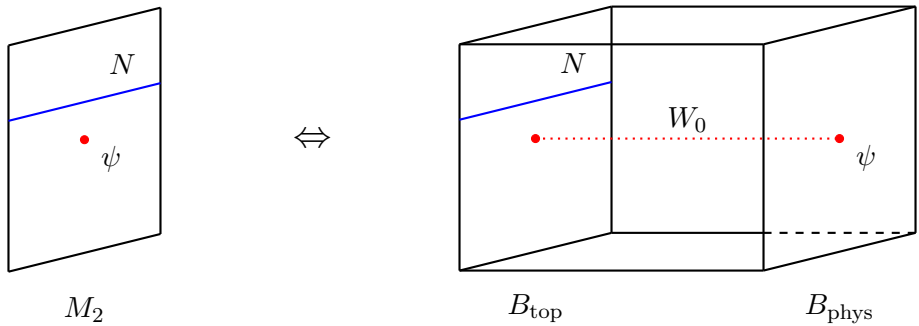

By definition, the local operators in $\mathcal{A}_{\mathcal{F}}(M_{1})$ are the ones invariant under the actions of all $N\in\mathcal{F}$ on $B_{\textrm{top}}$. From the SymTFT perspective, such an uncharged local operator is always represented by the identity line $W_0$ connecting the operator to $B_{\textrm{top}}$ as shown in Figure~\ref{Fig-Symmetric-Local-Operator}. Equivalently, we say that the endpoint of $W_0$ at $B_{\textrm{top}}$ furnishes a trivial representation of $\mathcal{F}$. Hence, an uncharged local operator is essentially trivial from the SymTFT perspective.

\subsubsection*{Twist operators}

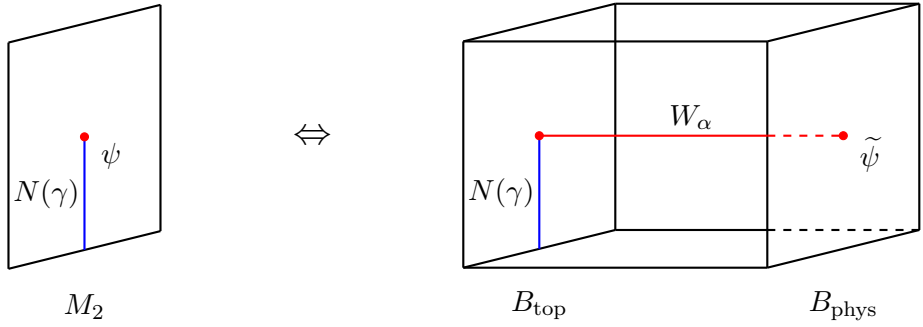
\begin{figure}[!h]
    \begin{equation}
            \begin{gathered}
        \begin{tikzpicture}
            \draw[thick] (0,0)--(0,3);
            \draw[thick] (0,3)--(2,3.5);
            \draw[thick] (2,3.5)--(2,0.5);
            \draw[thick] (2,0.5)--(0,0);
            % \draw[thick,red] (1,1.75)--(1,1.75+1.5);
            \draw[blue,thick] (1,1.75)--(1,0.25);
            \node at (0.5,1) {$N(\gamma)$};
            \filldraw[red] (1,1.75) circle (1.5pt); 
            % \draw[thick,blue] (1,1.75) ellipse (0.5 and 1);
            % \node at (1.5,3) {$\hat{W}$};
            % \node at (1.5,1.75+1) {$U(g)$};
            \node at (1.35,1.5) {$\psi$};
            % \draw[thick,blue] (0,2)--(2,2.5);
            \node at (1,-0.5) {$M_2$};
            \node at (4,1.75) {\Large $\Leftrightarrow$};
        \end{tikzpicture}
    \end{gathered} \qquad \qquad
            \begin{gathered}
        \begin{tikzpicture}
            \draw[thick] (0,0)--(0,3);
            \draw[thick] (0,3)--(2,3.5);
            \draw[thick] (2,3.5)--(2,0.5);
            \draw[thick] (2,0.5)--(0,0);
            \draw[thick] (4,0)--(4,3);
            \draw[thick] (4,3)--(6,3.5);
            \draw[thick] (6,3.5)--(6,0.5);
            \draw[thick] (6,0.5)--(4,0);
            \draw[thick] (0,0)--(4,0);
            \draw[thick] (0,3)--(4,3);
            \draw[thick] (2,3.5)--(6,3.5);
            \draw[thick] (2,0.5)--(4,0.5);
            \draw[thick,dashed] (4,0.5)--(6,0.5);
            \draw[thick,red] (1,1.75)--(4,1.75);
            \draw[thick,red,dashed] (4,1.75)--(5,1.75);
            \draw[blue,thick] (1,1.75)--(1,0.25);
            \node at (0.5,1) {$N(\gamma)$};
            \filldraw[red] (1,1.75) circle (1.5pt);
            \filldraw[red] (5,1.75) circle (1.5pt); 
            % \draw[thick,blue] (1,1.75) ellipse (0.5 and 1);
            \node at (3,2) {$W_{\alpha}$};
            % \draw[thick,blue] (0,2)--(2,2.5);
            % \node at (1.5,2.75) {$N$};
            \node at (5.35,1.5) {$\widetilde{\psi}$};
            \node at (1,-0.5) {$B_{\textrm{top}}$};
            \node at (5,-0.5) {$B_{\textrm{phys}}$};
            % \node at (1.5,3) {$\hat{W}$};
            % \node at (1.5,1.75+1) {$U(g)$};
        \end{tikzpicture}
    \end{gathered}\nonumber 
    \end{equation}
    \caption{A twist operator in the SymTFT picture. For the configuration on the right we have $N(\gamma)\in t(W_\alpha)$.}
    \label{Fig-Twist-Operator}
\end{figure}

For a given $\mathcal{L}$, a line $W_\alpha$ is not terminable on the $B_{\textrm{top}}$ if $L_{\alpha}=n_\alpha W_\alpha \notin \mathcal{L}$, $\forall n_\alpha > 0$~\cite{Bhardwaj:2023ayw}. Rather, a ``tail'' labeled by $N\in\mathcal{F}$ along $\gamma$ on $B_{\textrm{top}}$ going to infinity must be attached to such $W_\alpha$ as shown in Figure~\ref{Fig-Twist-Operator}. Such an operator $\psi N(\gamma)$ in $\mathcal{T}_\mathcal{F}$ on $M_2$, or its SymTFT incarnation $\widetilde{\psi} W_\alpha N(\gamma)$ in $M_2\times [0,1]$, is called a \emph{twist operator}.
Since the boundary condition at infinity is changed due to $N$, a twist operator should actually be thought of as an operator of a different theory characterized by the boundary condition at infinity prescribed by $N$. Therefore, all twist operators with $N(\gamma)$ stretching to infinity must be excluded from the von Neumann algebra $\mathcal{A}(M_{1})$ of $\mathcal{T}_{\mathcal{F}}$, since it is simply not an operator acting on the Hilbert space of $\mathcal{T}_{\mathcal{F}}$.

Despite the fact that twist operators with $N(\gamma)$ extending to infinity are not in $\mathcal{A}(M_{1})$, they are still very important because we can truncate $N(\gamma)$ and dress it at the point of truncation in a consistent manner as discussed in section~\ref{sec:Symm_vNA}. Before discussing such truncation from the SymTFT perspective, we note that it is possible that a line in the bulk extends to inequivalent tails on $B_{\textrm{top}}$. Conversely, the same tail can also be attached to different lines in the bulk. For a line $W_{\alpha}$ that can extend to different tails in $B_{\textrm{top}}$, we write
\begin{equation}\label{eq:W_all_tails}
    t(W_{\alpha}) = \bigoplus_{N_i\in\mathcal{F}} a_i N_i\,,\quad (a_i\in\mathbb{Z}_{\geq 0})
\end{equation}
where each $N_i$ (more precisely, $N_i(\gamma)$) is a tail to which $W$ can attach, and the physical meaning of $a_i$ will be explained in a moment. We note that in the notation~(\ref{eq:W_all_tails}), $L\in\mathcal{L}$ if the identity of $\mathcal{F}$, or a multiple of it, is a summand in $t(L)$. For $W_\alpha$ with $t(W_\alpha) = aN$ and $W_\beta$ with $t(W_\beta) = bM$, we have
\begin{equation}\label{eq:counting_Btop}
    \text{Hom}_\mathcal{F}(t(W_\alpha), t(W_\beta)) = \begin{cases}
        &\mathbb{C}^{ab}\,,\quad (N = M) \\
        &\emptyset\,,\quad (N\neq M)
    \end{cases} 
\end{equation}
where $ab$ is number of ways that $t(W_\alpha)$ can be joined to $t(W_\beta)$ when $N = M$. To see why, we first consider two tails $aN$ and $N$ on $B_{\textrm{top}}$. The coefficient $a$ means there is an $a$-fold ambiguity for choosing an $N$ on $B_{\textrm{top}}$, which in turn implies that there are $a$ inequivalent ways to join $aN$ with $N$~\footnote{The coefficient in~(\ref{eq:termitable_line}) is interpreted similarly.}. Now it should be clear that for $W_\alpha$ with tail $aN$ and $W_\beta$ with tail $bN$, there are $ab$ inequivalent ways to join them. Therefore, for generic $W_\alpha$ with tail $\bigoplus a_iN_i$ and $W_\beta$ with tail $\bigoplus b_iN_i$, we have
\begin{equation}\label{eq:Hom_Btop}
    \text{Hom}_\mathcal{F}(t(W_\alpha), t(W_\beta)) = \bigoplus_{i} \mathbb{C}^{a_ib_i}\,,
\end{equation}
with total dimension $\sum_i a_ib_i$. Physically, $\sum_i a_ib_i$ is the total number of inequivalent ways that $t(W_\alpha)$ can be joined with $t(W_\beta)$ on $B_{\textrm{top}}$, where each $a_ib_i$ is the number of ways that $i^{\textrm{th}}$ component of the two tails can be joined.

Physically, for $t(W_{\alpha}) = \bigoplus_i a_iN_i$ and $t(W_{\beta}) = \bigoplus_i b_iN_i$, when $a_ib_i$ is non-zero for certain $i$, $W_{\alpha}$ can be joined to $W_{\beta}$ through $N_i$. Hence, to check if $W_{\alpha}$ and $W_{\beta}$ can be joined on $B_{\textrm{top}}$, we need to look for the corresponding tails represented in terms of the elements of $\mathcal{F}$. As~(\ref{eq:counting_Btop}) counts the number of inequivalent ways the tails can be joined on $B_{\textrm{top}}$, it is natural to look for its SymTFT version in terms of $W$'s. For this, we have~\cite{Kong:2013aya, Cong:2016ayp, Bhardwaj:2023idu}
\begin{equation}\label{homorphism-coefficient}
    \text{Hom}_\mathcal{F}(t(W_\alpha), t(W_\beta)) = \text{Hom}_\text{SymTFT}(W_{\alpha},W_{\beta}\otimes \mathcal{L}) \,.
\end{equation}
Physically, while the LHS counts of the number of ways that two tails can be joined on $B_{\textrm{top}}$, the RHS of the above equation simply counts the times of appearances of $W_{\alpha}$ in the fusion $W_{\beta}\otimes \mathcal{L}$, and the symmetry $\mathcal{F}$ is implicitly determined by the choice of $\mathcal{L}$.

A terminable line $W_{\alpha} \in \mathcal{L}$ on $B_{\textrm{top}}$ (with its endpoint $\widetilde{\psi}(x)$ at $B_{\textrm{phys}}$) may become a twist operator $\widetilde{\psi}W_\alpha N(\gamma)$ under the action of an element of $\mathcal{F}$ as illustrated in Figure~\ref{Fig-SymTFT}. When this is the case, one says that $\psi(x)$ together with $\psi(x)N(\gamma)$ for all possible $N$'s which can be obtained via symmetry actions in $\mathcal{F}$ form a multiplet in $\mathcal{T}_{\mathcal{F}}$.

\subsubsection*{Bi-twist and bi-local operators}

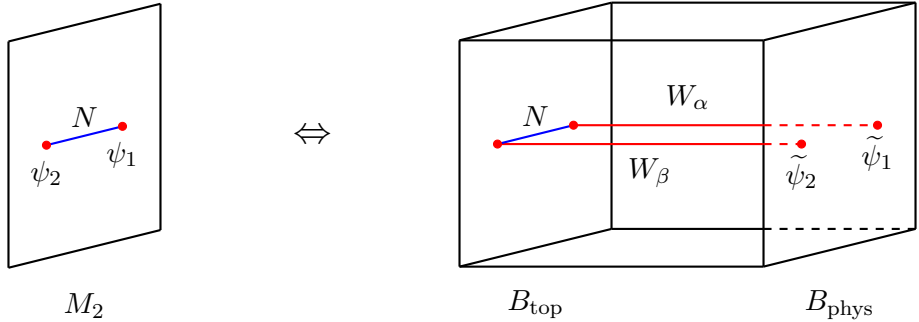
\begin{figure}[!h]
    \begin{equation}
        \begin{gathered}
        \begin{tikzpicture}
            \draw[thick] (0,0)--(0,3);
            \draw[thick] (0,3)--(2,3.5);
            \draw[thick] (2,3.5)--(2,0.5);
            \draw[thick] (2,0.5)--(0,0);
            % \draw[thick,red] (1,1.75)--(1,1.75+1.5);
            % \draw[blue,thick] (1,1.75)--(1,0.25);
            \node at (1,2) {$N$};
            \draw[blue,thick] (0.5,1.625)--(1.5,1.875);
            \filldraw[red] (0.5,1.625) circle (1.5pt); 
            \filldraw[red] (1.5,1.875) circle (1.5pt); 
            \node at (1.5,1.5) {$\psi_1$};
            \node at (0.5,1.25) {$\psi_2$};
            % \draw[thick,blue] (0,2)--(2,2.5);
            \node at (1,-0.5) {$M_2$};
            \node at (4,1.75) {\Large $\Leftrightarrow$};
        \end{tikzpicture}
    \end{gathered} \qquad \qquad
            \begin{gathered}
        \begin{tikzpicture}
            \draw[thick] (0,0)--(0,3);
            \draw[thick] (0,3)--(2,3.5);
            \draw[thick] (2,3.5)--(2,0.5);
            \draw[thick] (2,0.5)--(0,0);
            \draw[thick] (4,0)--(4,3);
            \draw[thick] (4,3)--(6,3.5);
            \draw[thick] (6,3.5)--(6,0.5);
            \draw[thick] (6,0.5)--(4,0);
            \draw[thick] (0,0)--(4,0);
            \draw[thick] (0,3)--(4,3);
            \draw[thick] (2,3.5)--(6,3.5);
            \draw[thick] (2,0.5)--(4,0.5);
            \draw[thick,dashed] (4,0.5)--(6,0.5);
            \draw[thick,red] (1.5,1.875)--(4,1.875);
            \draw[thick,red,dashed] (4,1.875)--(5.5,1.875);
            \draw[thick,red] (0.5,1.625)--(4,1.625);
            \draw[thick,red,dashed] (4,1.625)--(4.5,1.625);
            \node at (1,2) {$N$};
            \draw[blue,thick] (0.5,1.625)--(1.5,1.875);
            \filldraw[red] (0.5,1.625) circle (1.5pt); 
            \filldraw[red] (1.5,1.875) circle (1.5pt); 
            \filldraw[red] (5.5,1.875) circle (1.5pt); 
            \filldraw[red] (4.5,1.625) circle (1.5pt); 
            % \draw[thick,blue] (1,1.75) ellipse (0.5 and 1);
            \node at (3,2.25) {$W_{\alpha}$};
            \node at (2.5,1.25) {$W_{\beta}$};
            % \draw[thick,blue] (0,2)--(2,2.5);
            % \node at (1.5,2.75) {$N$};
            \node at (5.5,1.5) {$\widetilde{\psi}_1$};
            \node at (4.5,1.25) {$\widetilde{\psi}_2$};
            \node at (1,-0.5) {$B_{\textrm{top}}$};
            \node at (5,-0.5) {$B_{\textrm{phys}}$};
            % \node at (1.5,3) {$\hat{W}$};
            % \node at (1.5,1.75+1) {$U(g)$};
        \end{tikzpicture}
    \end{gathered}\nonumber 
    \end{equation}
    \caption{The bi-twist operator in the SymTFT picture. When $N=1$ is the identity line, the operator becomes a bi-local operator in $\mathcal{T}_{\mathcal{F}}$.}
    \label{Fig-bitwist-Operator}
\end{figure}

Let us consider two lines $W_{\alpha}$ and $W_{\beta}$ in the bulk. As discussed previously, they can be connected on $B_{\textrm{top}}$ by a line $N$ iff
\begin{equation}\label{eq:bi-twist_condition}
    \text{Hom}_{\textrm{SymTFT}}(W_\alpha,W_\beta\times\mathcal{L}) \neq \emptyset\,.
\end{equation}
Therefore, when~(\ref{eq:bi-twist_condition}) holds, $W_\alpha$ and $W_\beta$ together with the line connecting them on $B_{\textrm{top}}$ can form a bi-twist operator as shown in Figure~\ref{Fig-bitwist-Operator}.
In particular, when $N=1$, or equivalently when $W_{\alpha}, W_{\beta} \in \mathcal{L}$, the endpoints of $W_\alpha$ and $W_\beta$ form a bi-local operator in 2D. Similar to a single local operator, a generic bi-local or bi-twist operator can become a different bi-local or bi-twist operator under a symmetry action (cf.~Figure~\ref{Fig-SymTFT}). Since both the bi-twist and bi-local operators have finite volume support and do not change the boundary condition at infinity, we choose to include them in $\mathcal{A}(M_{1})$ and the same choice has been made in~\cite{Shao:2025mfj}.

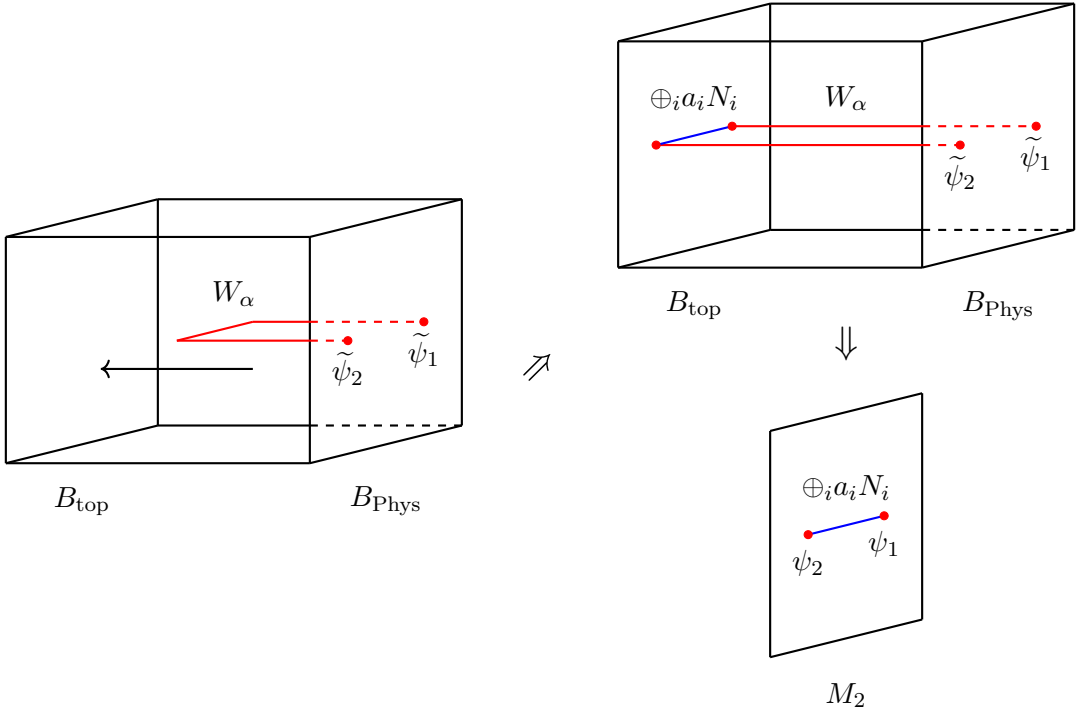
\begin{figure}[h]
    \begin{equation*}
            \begin{gathered}
        \begin{tikzpicture}
            \draw[thick] (0,0)--(0,3);
            \draw[thick] (0,3)--(2,3.5);
            \draw[thick] (2,3.5)--(2,0.5);
            \draw[thick] (2,0.5)--(0,0);
            \draw[thick] (4,0)--(4,3);
            \draw[thick] (4,3)--(6,3.5);
            \draw[thick] (6,3.5)--(6,0.5);
            \draw[thick] (6,0.5)--(4,0);
            \draw[thick] (0,0)--(4,0);
            \draw[thick] (0,3)--(4,3);
            \draw[thick] (2,3.5)--(6,3.5);
            \draw[thick] (2,0.5)--(4,0.5);
            \draw[thick,dashed] (4,0.5)--(6,0.5);
            \draw[thick,red] (3.25,1.875)--(4,1.875);
            \draw[thick,red,dashed] (4,1.875)--(5.5,1.875);
            \draw[thick,red] (2.25,1.625)--(4,1.625);
            \draw[thick,red,dashed] (4,1.625)--(4.5,1.625);
            % \node at (1,2) {$L$};
            % \draw[blue,thick] (0.5,1.625)--(1.5,1.875);
            % \filldraw[red] (0.5,1.625) circle (1.5pt); 
            % \filldraw[red] (1.5,1.875) circle (1.5pt); 
            \filldraw[red] (5.5,1.875) circle (1.5pt); 
            \filldraw[red] (4.5,1.625) circle (1.5pt); 
            \draw[red,thick] (2.25,1.625)--(3.25,1.875);
            % \draw[thick,blue] (1,1.75) ellipse (0.5 and 1);
            \begin{scope}[thick,decoration={markings,mark=at position 1 with {\arrow{>}}}] 
            \draw[postaction={decorate}] (3.25,1.25)--(1.25,1.25);
            \end{scope}
            \node at (3,2.25) {$W_{\alpha}$};
            % \draw[thick,blue] (0,2)--(2,2.5);
            % \node at (1.5,2.75) {$N$};
            \node at (5.5,1.5) {$\widetilde{\psi}_1$};
            \node at (4.5,1.25) {$\widetilde{\psi}_2$};
            \node at (1,-0.5) {$B_{\textrm{top}}$};
            \node at (5,-0.5) {$B_{\textrm{Phys}}$};
            % \node at (1.5,3) {$\hat{W}$};
            % \node at (1.5,1.75+1) {$U(g)$};
        \end{tikzpicture}
    \end{gathered} \qquad \textrm{\Large $\rotatebox[origin=c]{45}{\(\Rightarrow\)}$} \qquad
    \begin{gathered}
        \begin{tikzpicture}
            \draw[thick] (0,0)--(0,3);
            \draw[thick] (0,3)--(2,3.5);
            \draw[thick] (2,3.5)--(2,0.5);
            \draw[thick] (2,0.5)--(0,0);
            \draw[thick] (4,0)--(4,3);
            \draw[thick] (4,3)--(6,3.5);
            \draw[thick] (6,3.5)--(6,0.5);
            \draw[thick] (6,0.5)--(4,0);
            \draw[thick] (0,0)--(4,0);
            \draw[thick] (0,3)--(4,3);
            \draw[thick] (2,3.5)--(6,3.5);
            \draw[thick] (2,0.5)--(4,0.5);
            \draw[thick,dashed] (4,0.5)--(6,0.5);
            \draw[thick,red] (1.5,1.875)--(4,1.875);
            \draw[thick,red,dashed] (4,1.875)--(5.5,1.875);
            \draw[thick,red] (0.5,1.625)--(4,1.625);
            \draw[thick,red,dashed] (4,1.625)--(4.5,1.625);
            \node at (1,2.25) {$\oplus_i a_i N_i$};
            \draw[blue,thick] (0.5,1.625)--(1.5,1.875);
            \filldraw[red] (0.5,1.625) circle (1.5pt); 
            \filldraw[red] (1.5,1.875) circle (1.5pt); 
            \filldraw[red] (5.5,1.875) circle (1.5pt); 
            \filldraw[red] (4.5,1.625) circle (1.5pt); 
            % \draw[thick,blue] (1,1.75) ellipse (0.5 and 1);
            \node at (3,2.25) {$W_{\alpha}$};
            % \node at (2.5,1.25) {$W_{\bar{\alpha}}$};
            % \draw[thick,blue] (0,2)--(2,2.5);
            % \node at (1.5,2.75) {$N$};
            \node at (5.5,1.5) {$\widetilde{\psi}_1$};
            \node at (4.5,1.25) {$\widetilde{\psi}_2$};
            \node at (1,-0.5) {$B_{\textrm{top}}$};
            \node at (5,-0.5) {$B_{\textrm{Phys}}$};
            \node at (3,-1) {\Large $\Downarrow$};
            % \node at (1.5,3) {$\hat{W}$};
            % \node at (1.5,1.75+1) {$U(g)$};
        \end{tikzpicture}\\
        \begin{tikzpicture}
            \draw[thick] (0,0)--(0,3);
            \draw[thick] (0,3)--(2,3.5);
            \draw[thick] (2,3.5)--(2,0.5);
            \draw[thick] (2,0.5)--(0,0);
            % \draw[thick,red] (1,1.75)--(1,1.75+1.5);
            % \draw[blue,thick] (1,1.75)--(1,0.25);
            \node at (1,2.25) {$\oplus_i a_i N_i$};
            \draw[blue,thick] (0.5,1.625)--(1.5,1.875);
            \filldraw[red] (0.5,1.625) circle (1.5pt); 
            \filldraw[red] (1.5,1.875) circle (1.5pt); 
            \node at (1.5,1.5) {$\psi_1$};
            \node at (0.5,1.25) {$\psi_2$};
            % \draw[thick,blue] (0,2)--(2,2.5);
            \node at (1,-0.5) {$M_2$};
        \end{tikzpicture}
    \end{gathered}
    \end{equation*}
    \caption{A set of symmetric bi-twist (or bi-local depending on whether $\oplus_{i} a_i N_i \propto 1$) operators can be constructed by bending the bulk operator $W_{\alpha}$ and dragging it onto $B_{\textrm{top}}$. }
    \label{Fig-patch-Operator}
\end{figure}

We now look for bi-twist and bi-local operators that are in $\mathcal{A}_{\mathcal{F}}(M_{1})$. The construction of such operators, which was a bit cumbersome and subject to many edge cases in section~\ref{sec:Symm_vNA} from the perspective of the physical QFT living on the boundary, becomes transparent from the SymTFT perspective in the bulk. We consider connecting $\widetilde{\psi}_1$ with $\widetilde{\psi}_2$ by a line $W_\alpha$ in the bulk, assuming such a line exists. Then aftering dragging and projecting $W_\alpha$ onto $B_{\textrm{top}}$, the composite operator $\widetilde{\psi}_1W_\alpha\widetilde{\psi}_2$ must commute with all $\mathcal{F}$ on $B_{\textrm{top}}$, since we can simply drag it back into the bulk to make it not touch anything on $B_{\textrm{top}}$. The operators are shown in Figure~\ref{Fig-patch-Operator} and are referred to as patch operators~\cite{Ji:2019jhk,Wu:2020yxa,Chatterjee:2022kxb,Chatterjee:2022jll,Inamura:2023ldn}. 

Recall that when $W_\alpha$ can be connected to different tails $N_i$ on $B_{\textrm{top}}$, we write $t(W_\alpha) = \bigoplus a_i N_i$ where $a_i$'s specify the ambiguities for each $N_i$. This means one can connect $W_\alpha$ to $W_{\overline{\alpha}}$ (the overline indicates orientation-reversal) by an $N_i$ on $B_{\textrm{top}}$. However, one generically cannot drag $W_\alpha N_i W_{\overline{\alpha}}$ into the bulk in the manner discussed previously. Otherwise, $W_\alpha N_i W_{\overline{\alpha}}$ can be dragged into the bulk then dragged back onto $B_{\textrm{top}}$ with a different $N_j$ connecting them where the transformation from $N_i$ to $N_j$ is realized by a symmetry action in $\mathcal{F}$. Hence $W_\alpha N_i W_{\overline{\alpha}}$ is non-invariant under an action in $\mathcal{F}$, which contradicts the fact that any bi-twist operator which can be freely dragged into the bulk must commute with all actions in $\mathcal{F}$. Therefore, to ensure that a boundary line can be freely dragged into the bulk, we must connect $W_\alpha$ and $W_{\overline{\alpha}}$ by the full $\bigoplus_i a_i N_i$ as illustrated in Figure~\ref{Fig-patch-Operator}. In other words, an uncharged bi-twist/bi-local operator is realized by the operator $W_\alpha \left( 
\bigoplus_i a_i N_i \right) W_{\overline{\alpha}}$ in the SymTFT in the bulk. These operators are therefore by definition in $\mathcal{A}_\mathcal{F}(M_{1})$ and will play important roles in section~\ref{sec:Add_Haag_Violation}.

\subsection{Generalization to Higher Dimensions}\label{sec:higher-dimensions}

In this section, we generalize the above construction to higher dimensions, where the distinction between bi-local operators and bi-twist operators becomes manifest. While the bi-local and the bi-twist operators can be constructed in essentially the same manner as illustrated in Figure~\ref{Fig-patch-Operator}, one has to be careful about the dimensionality of various operators. For example, when $\mathcal{F}$ is a $0$-form symmetry, while an uncharged bi-local can still be constructed by a line in the bulk ending on $B_{\textrm{top}}$, a bi-twist operator should be viewed as a $(d-1)$-dimensional hypersurface in the bulk joined with a $(d-1)$-dimensional tail on $B_{\textrm{top}}$.

The $BF$-SymTFT proposed in~\cite{Jia:2025jmn} provides a natural framework for the construction of such operators in general dimension, at least for invertible symmetries. To be concrete, we focus on QFT $\mathcal{T}_{SU(n)}$ with non-anomalous $SU(n)$ flavor symmetry on a $d$-dimensional manifold $M_d$. The SymTFT action in the bulk is an $SU(n)$ $BF$-action
\begin{equation}
    S = \int_{M_{d+1}} \text{tr} B\wedge F\,,
\end{equation}
for $M_{d+1}\cong M_d\times [0,1]$ and the QFT lives on $B_{\textrm{phys}} \cong M_d\times \{1\}$. To construct a bi-local operator from the SymTFT perspective, we consider a Wilson line $W_{\mathbf{r}}$ connecting two local operators on $B_{\textrm{phys}}$
\begin{equation}
    \tr\widetilde{\psi}_1(x) W_{\mathbf{r}}(\gamma_{xy}) \widetilde{\psi}_2(y)\equiv \widetilde{\psi}_1(x) \mathcal{P} \exp \left( i \int_{\gamma_{xy}} A \right)\widetilde{\psi}_{2}(y) \,,
\end{equation}
for a path $\gamma_{xy}$ connecting $x$ and $y$ in the bulk. Since $\mathcal{P} \exp \left( i \int_{\gamma_{xy}} A \right)$ with fixed $x$ and $y$ is topological in $M_{d+1}$, it can be freely restricted onto or dragged away from $B_{\textrm{top}}$, therefore such an operator must become an uncharged bi-local operator $\tr(\psi_1(x)\psi_2(y))$ after shrinking the interval (cf.~(\ref{eq:bi-local_ops})).

The construction of a bi-twist operator with codimension-1 support is a bit more tricky. The bulk topological operator in this case is $U_{[\alpha]}(\Sigma_{d-1})$ with codimension-2 support $\Sigma_{d-1}$ in the bulk, which depends on a conjugacy class $[\alpha]$ of $SU(n)$. Without reviewing the details of~\cite{Jia:2025jmn}, the key observation is that a (hyper)surface operator of this type that depends on a single element of $SU(n)$ cannot be topological and hence cannot be freely dragged into the bulk. Therefore, unless $\alpha$ takes certain special values, e.g. $\alpha \in Z(SU(n))$ which is the center element of $SU(n)$, the tail of $U_{[\alpha]}(\Sigma_{d-1})$ on $B_{\textrm{top}}$ cannot commute with all elements of $SU(n)$. Note that there is no contradiction here since any $\alpha \in Z(G)$ is the only representative of its own conjugacy class. However, when $\alpha\notin Z(SU(n))$, one has to include all elements in its conjugacy class to construct the topological operator $U_{[\alpha]}(\Sigma_{d-1})$ in order for it to be freely moved away from $B_{\textrm{top}}$ hence is a candidate in $\mathcal{A}_{SU(n)}(M_{d-1})$.

More precisely, this topological operator can be written explicitly as~\cite{Jia:2025jmn}
\begin{equation}
    U_{[\alpha]}(\Sigma_{d-1}) = \int dg \exp\left( i\int_{\Sigma_{d-1}}(\widetilde{\alpha}, B) \right)\,,
\end{equation}
where $dg$ is the Haar measure of $G$ and $\widetilde{\alpha}$ is the parallel transport of $g\alpha(p)g^{-1}$ for a covariantly constant $\alpha$ at an arbitrary $p\in \Sigma_{d-1}$ by the holonomy of $A$. One may recognize that this is a concrete realization of~(\ref{eq:W_all_tails}), where the integral plays the role of the direct sum in~(\ref{eq:W_all_tails}). We then terminate such $U_{[\alpha]}(\Sigma_{d-1})$ on $B_{\textrm{phys}}$ with suitable $\widetilde{\Psi}_1$ and $\widetilde{\Psi}_2$ at the two boundaries of $\Sigma_{d-1}$ to obtain a bi-twist operator $\tr \Psi_1 N(\Sigma_{d-1}) \Psi_2 \in \mathcal{A}_{SU(n)}(M_{d-1})$ after projecting to $\mathcal{T}_{SU(n)}$ (cf.~Figure~\ref{fig:TwistOperator}). In this work, we will not explicitly give a full construction of bi-twist operators in the presence of non-invertible symmetries; rather, we just note that the construction should be similar to the one given above.

In the above discussion, we assumed the topological boundary $B_{\textrm{top}}$ supports the $0$-form $SU(n)$ flavor symmetry. If we gauge the non-anomalous $SU(n)$ flavor symmetry\footnote{The $SU(n)$ symmetry is gauged by summing over flat $SU(n)$-connections on $M_{d}$.} in $\mathcal{T}_{SU(n)}$, a $(d-2)$-form $\textrm{Rep}(SU(n))$ non-invertible symmetry whose charged objects are $(d-2)$-dimensional extended operator will emerge. In the SymTFT picture, the gauging corresponds to switching to another topological boundary $B'_{\textrm{top}}$ with boundary condition such that only $(d-1)$-dimensional extended operator $U_{[\alpha]}$ can end on $B'_{\textrm{top}}$. The Wilson line operators $W$ in the bulk will become the generators $(d-2)$-form $\textrm{Rep}(SU(n))$ symmetry on $B'_{\textrm{top}}$. The roles of bi-local operators and bi-twist operators are exchanged after the gauging. The extended operator $U_{[\alpha]}(\Sigma_{d-1})$ terminated at $\widetilde{\Psi}_1$ and $\widetilde{\Psi}_2$ on $B_{\textrm{phys}}$ become a bi-local operator $\tr\widetilde{\Psi}_1\widetilde{\Psi}_2\in\mathcal{A}_{\textrm{Rep}(SU(n))}(M_{d-1})$ and is uncharged under the $(d-2)$-form symmetry after projecting to $\mathcal{T}_{\textrm{Rep}(SU(n))}$. On the other hand, the local operators $\widetilde{\psi}_1(x), \widetilde{\psi}_2(y)$ on $B_{\textrm{phys}}$ connected by a bulk Wilson line operators $W_{\textbf{r}}(\gamma_{xy})$ will project to the bi-twist operator $\psi_1(x) \mathcal{P} \exp \left( i \int_{\gamma_{xy}} A \right)\psi_2(y) \in \mathcal{A}_{\textrm{Rep}(SU(n))}(M_{d-1})$ after projecting to $\mathcal{T}_{\textrm{Rep}(SU(n))}$.

We can also gauge only the center subgroup $Z(SU(n)) = \mathbb{Z}_n$ in $\mathcal{T}_{SU(n)}$, the resulting symmetry would be a mixture between a 0-form $PSU(n)$ symmetry and a $(d-2)$-form  $\mathbb{Z}_n$ symmetry. In the SymTFT picture, only the Wilson line operators $W_{\textbf{r}}$ whose representation $\mathbf{r}$ is neutral under the $\mathbb{Z}_N$ center can end on the new topological boundary. They give rise to the local operators $\psi(x)$ charged under the $PSU(n)$ symmetry after projecting to $\mathcal{T}_{PSU(n)\times \mathbb{Z}^{(d-2)}_n}$. The $(d-1)$-dimensional extended operator $U_{[\alpha]}$ with the conjugacy class $[\alpha] \in Z(SU(n))$ can also end on the topological boundary, and they give rise to the $(d-2)$-dimensional extended operators $\Psi(\Sigma_{d-2})$ charged under the $(d-2)$-form $\mathbb{Z}_n$ symmetry. Therefore, we have two kinds of symmetric bi-local (extended) operators in $\mathcal{T}_{PSU(n)\times \mathbb{Z}^{(d-2)}_n}$. Other Wilson line operators $W'$ and extended operators $U'$ cannot end on the topological boundary, and they should connect to $1$-dimensional or $(d-1)$-dimensional tails. The tails are generators of the $(d-2)$-form $\mathbb{Z}_n$ symmetry and $0$-form $PSU(n)$ symmetry on the topological boundary. As a consequence, we also have two kinds of symmetric bi-twist (extended) operators in $\mathcal{T}_{PSU(n)\times \mathbb{Z}^{(d-2)}_n}$ after such $\mathbb{Z}_n$-gauging.

In general, one can study a $d$-dimensional theory  $\mathcal{T}_{\mathcal{F}}$ whose $\mathcal{F}$ contains $0$-form and/or $(d-2)$-form symmetries. In the $(d+1)$-dimensional SymTFT, we can define the topological boundary $B_{\textrm{top}}$ via choosing the collection of line operators $W_{\alpha}$ and $(d-1)$-dimensional extended operator $U_{\mu}$ in the bulk that can end simultaneously on $B_{\textrm{top}}$ and write formally as (cf.~\eqref{eq:Lagrangian_algebra})
    \begin{equation}
        \mathcal{L} = \left(\bigoplus_\alpha W_{\alpha}\right)  \bigoplus \left(\bigoplus_{\mu} U_{\mu} \right) \,.
    \end{equation}
If $\mathcal{F}$ happens to have other $p$-form symmetries in it, one must also include other compatible endable extended operators in the SymTFT to $\mathcal{L}$.

As a final remark, note that we see a $(d-2)$-form symmetry arise as the consequence of gauging 0-form symmetry $\mathcal{F}$. On the other hand, if $\mathcal{F}$ is a pure $0$-form symmetry, then the local operators are not allowed to transform to twist operators under non-invertible symmetry action when $d>2$, since the tail of the twist operators are $1$-dimensional which implies the $(d-2)$-form symmetry should coexist in the theory. In the following discussion, we will allow the presence of $(d-2)$-form symmetry in $\mathcal{F}$, which can interact non-trivially with the $0$-form symmetry to allow the transition from local operators to twist operators. We will denote the generator of $(d-2)$-form symmetry by $\eta(\gamma)$ for a line $\gamma$, the fact that it can attach to a non-invertible 0-form generator $N(M_{d-1})$ supported on $M_{d-1}$ implies 
\begin{equation}
    \eta(\gamma) N(M_{d-1}) = N(M_{d-1}) \eta(\gamma) = N(M_{d-1})\,,\quad (\gamma \subset M_{d-1})
\end{equation}
such that $\eta(\gamma)$ can be absorbed into $N(M_{d-1})$. Multiply $\overline{N}(M_{d-1})$ on both side and assume $1 \in N\overline{N}$, one has $\eta(\gamma) \in N(M_{d-1})\overline{N}(M_{d-1})$ for any $\gamma \subset M_{d-1}$. Therefore, one would expect
    \begin{equation}\label{Condensation-defect}
        N(M_{d-1}) \overline{N} (M_{d-1}) = \mathcal{C}(M_{d-1})\,,
    \end{equation}
where $\mathcal{C}(M_{d-1})$ is known as a condensation of $\eta(\gamma)$ on $M_{d-1}$. Notice that \eqref{Condensation-defect} does not contradict with $1\in N \overline{N}$ since the identity element $1$ is also an ingredient of $\mathcal{C}$. For example, when $M_{d-1}$ is a compact manifold and $\eta(\gamma)$ generates an $\mathbb{Z}_k$ $(d-2)$-form symmetry, we can write
    \begin{equation}
        \mathcal{C}(M_{d-1}) = \bigoplus_{\gamma \in H_1(M_{d-1},\mathbb{Z}_k)}  \eta (\gamma)\,,
    \end{equation}
which is known as $1$-gauging the $(d-3)$-form symmetry along $M_{d-1}$\cite{Roumpedakis:2022aik}.

In the literature, similar $N$-defects are constructed as the duality defects of the theory under the gauging of a $p$-form symmetry, and they are symmetries of the theory only when it is self-dual. For example, the $2d$ Ising model at temperature $T$ is mapped to itself with dual temperature $T'$ under the Kramers-Wannier (KW) transformation, or $\mathbb{Z}_2$-gauging. The KW duality defect $N$ becomes a symmetry only at the critical temperature $T_c$. Duality defects in $4d$ on gauging $1$-form symmetries are also studied in \cite{Kaidi:2021xfk,Choi:2021kmx,Choi:2022zal,Choi:2022jqy,Antinucci:2022vyk,Kaidi:2022cpf}. In the following, we will assume the existence of such $N$ as a non-invertible symmetry and not delve into the details of its construction.

\subsection{Relation to the Non-SymTFT Construction and a Sub-algebra of the Lagrangian Algebra}
\label{sec:Relation_OpSpec_SymTFT}

It should be clear from our construction in Section~\ref{sec:SvNA_OpSpec} that $\mathcal{A}_{\mathcal{F}}(M_{1})$ contains a bi-local operator whenever there exists a local operator $\psi(x)$ that does not become a twist operator connecting to a defect line under any non-intertible element of $\mathcal{F}$. The SymTFT construction presented in Section~\ref{sec:SvNA_SymTFT} makes this construction sharper by encapsulating the above somewhat complicated condition into a simpler one constraining the property of the Lagrangian algebra $\mathcal{L}$.

Recall that a charged local operator $\psi(x)$ corresponds to a terminable line $W\in\mathcal{L}$ with one end on $B_{\textrm{top}}$ and the other end on $\widetilde{\psi}(x)$. Requiring $W$ be terminable means $t(W) = a_0 1\oplus (\oplus_{i\neq 0} a_i N_i)$ with non-zero $a_0$. If $t(W) \neq a_0 1$, there must exist certain non-invertible symmetries, say $M\in\mathcal{F}$, upon which the identity tail $1$ can be transformed to another $N_i$ of $t(W)$. Therefore, after projection to $\mathcal{T}_{\mathcal{F}}$, $\psi(x)$ cannot be invariant under $M$ as it must become a twist operator joined to $N_i$ defect line. Therefore, in order to have a bi-local operator in $\mathcal{A}_\mathcal{F}(M_{1})$, we must require that there exists $L\in \mathcal{L}$ such that $t(L) \propto 1$. We define $\widetilde{\mathcal{L}} \subset \mathcal{L}$ to be the subset containing all $L\in\mathcal{L}$ such that $t(L) \propto 1$ on $B_{\textrm{top}}$. Therefore, the necessary and sufficient condition for $\mathcal{A}_\mathcal{F}(M_{1})$ containing a bi-local operator is $\widetilde{\mathcal{L}}\neq W_0$, which means it contains non-trivial elements other than the bulk identity.

On the other hand, there always exist symmetric bi-twist operators in $\mathcal{A}_{\mathcal{F}}(M_{1})$. Clearly, for any bulk line operator $W' \notin \widetilde{\mathcal{L}}$ with $t(W')\neq a_0 1$, it can be transformed into a twist operator connecting to a defect line on $B_{\textrm{top}}$. We then construct the lines $\widetilde{\psi}_1W'$ and $\widetilde{\psi}_2W'$ in $M_3$ and connect the lines by $t(W')=\oplus_ia'_iN_i$ on $B_{\textrm{top}}$. Since this operator can be freely dragged into $M_3$, it must be uncharged under all symmetry actions in $\mathcal{F}$. After projection to $\mathcal{T}_{\mathcal{F}}$, the two operators $\psi_1$ and $\psi_2$ connecting by $t(W')=\oplus_ia'_iN_i$ transform under two representations of $\mathcal{F}$ that dual to each other, since they are lifted to $W'$ and its orientation reversal in the bulk SymTFT.

In higher dimension, we similarly define $\widetilde{\mathcal{L}} \subset \mathcal{L}$ to be the subset containing all line operators $W$ and $(d-1)$-dimensional extended operators $U$ in $\mathcal{L}$ that cannot attach to non-trivial $1$-dimensional and $(d-1)$-dimensional tails on $B_{\textrm{top}}$. Then $\mathcal{A}_{\mathcal{F}}(M_{d-1})$ contains bi-local (extended) operators if $\widetilde{\mathcal{L}}\neq W_0\oplus U_0$, which means it contains operators other than the trivial line and $(d-1)$-dimensional extended operators. On the other hand, one can always construct symmetric bi-twist operators using line operators $W'\notin \widetilde{\mathcal{L}}$ or $(d-1)$-dimensional extended operators $U' \notin \widetilde{\mathcal{L}}$, connecting two local operators or two $(d-2)$-dimensional extended operators on $B_{\textrm{phys}}$ as discussed in Section~\ref{sec:higher-dimensions}.

In summary, we introduce $\widetilde{\mathcal{L}} \subset \mathcal{L}$ to be the subset generated by all line operators $W\in\mathcal{L}$ and $(d-1)$-dimensional extended operator $U\in\mathcal{L}$ that cannot join to any non-trivial tail on $B_{\textrm{top}}$. Thus, any local operator on $B_{\textrm{phys}}$ that is the endpoint of $W$ cannot be transformed to a twist operator under any symmetry action in $\mathcal{F}$. A similar argument holds for any $(d-2)$-dimensional extended operator on $B_{\textrm{phys}}$ that is the ending locus of $U$. Clearly, $\widetilde{\mathcal{L}}$ is closed under fusion.

\section{Additivity and Haag Duality in the Presence of Bi-local and Bi-twist Operators}
\label{sec:Add_Haag_Violation}

We are now ready to discuss the implications of the operator content of $\mathcal{A}_\mathcal{F}(M_{d-1})$ on certain properties of it. In particular, we focus on additivity and Haag duality of $\mathcal{A}_\mathcal{F}(M_{d-1})$. We anticipate the case where $\mathcal{F}$ is a mixture of $0$-form and $(d-2)$-form symmetry as discussed at the end of section~\ref{sec:higher-dimensions}, which allows the transition between local operators and twist operators under non-invertible symmetries. In general, we will consider four kinds of bi-local and bi-twist operators in $\mathcal{A}_\mathcal{F}(M_{d-1})$.
\begin{itemize}
    \item Bi-local operators $\textrm{tr} \psi_1(x) \psi_2(y)$;
    \item Bi-local extended operators $\textrm{tr}\Psi_1(\Sigma_{d-2}) \Psi_2(\Sigma'_{d-2})$;
    \item Bi-twist operators $\tr \psi_1(x) \eta(\gamma_{x y}) \psi_2(y)$;
    \item Bi-twist extended operators $\tr \Psi_1(\Sigma_{d-2}) N(\Sigma_{d-1}) \Psi_2(\Sigma'_{d-2})$ with $\partial \Sigma_{d-1} = \Sigma_{d-2} \bigcup \overline{\Sigma'_{d-2}}$.
\end{itemize}

Following~\cite{Shao:2025mfj}, we will show the following holds in general, regardless of the details of the dynamics of $\mathcal{T}_\mathcal{F}$:
\begin{itemize}
    \item Additivity of $\mathcal{A}_{\mathcal{F}}(M_{d-1})$ is violated whenever $\mathcal{A}_\mathcal{F}(M_{d-1})$ contains bi-local operators $\textrm{tr} \psi_1(x) \psi_2(y)$ or bi-local $(d-2)$-dimensional extended operators $\textrm{tr}\Psi_1(\Sigma_{d-2}) \Psi_2(\Sigma'_{d-2})$.
    \item Haag duality of $\mathcal{A}_{\mathcal{F}}(M_{d-1})$ is violated whenever $\mathcal{A}_\mathcal{F}(M_{d-1})$ contains bi-twist operators $\tr\Psi_1(\Sigma_{d-2})N(\Sigma_{d-1})\Psi_2(\Sigma'_{d-2})$ or $\tr \psi_1(x)\eta(\gamma_{xy}) \psi_2(y)$ that commute with all components of bi-local (extended) operators in $\mathcal{A}_{\mathcal{F}}(M_{d-1})$.
\end{itemize}
We will first show the above holds in general dimensions, then we will specialize to 2d, where the conditions in the above claim can be made concrete using the Lagrangian algebra explicitly.

\subsection{Violation of Additivity and Haag Duality in General Dimensions}
\label{sec:Add_Haag_Violation_Gen_Dim}

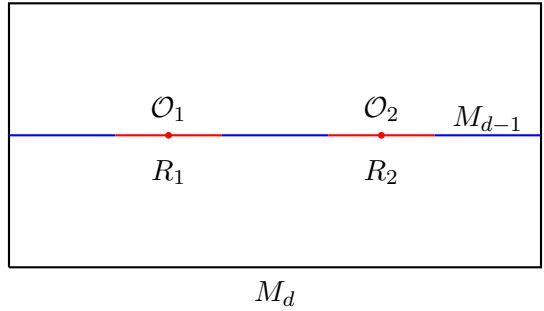
\begin{figure}[!h]
    \begin{equation}
            \begin{gathered}
        \begin{tikzpicture}[scale=0.7]
            \draw[thick] (0,0)--(0,5)--(10,5)--(10,0)--(0,0);
            \draw[thick,blue] (0,2.5)--(2,2.5);
            \draw[thick,red] (2,2.5)--(4,2.5);
            \draw[thick,blue] (4,2.5)--(6,2.5);
            \draw[thick,red] (6,2.5)--(8,2.5);
            \draw[thick,blue] (8,2.5)--(10,2.5);
            \filldraw[red] (3,2.5) circle (1.5pt); 
            \filldraw[red] (7,2.5) circle (1.5pt); 
            \node at (3,3) {$\mathcal{O}_1$};
            \node at (7,3) {$\mathcal{O}_2$};  
            \node at (3,1.8) {$R_1$};
            \node at (7,1.8) {$R_2$};  
            \node at (9,2.8) {$M_{d-1}$};
            \node at (5,-0.5) {$M_{d}$};
        \end{tikzpicture}
    \end{gathered} \nonumber
    \end{equation}
    \caption{Putting a bi-local operator on $M_{d-1}$. Here the operator $\mathcal{O}$ can be either local operator $\psi(x)$ or extended operator $\Psi(\Sigma_{d-2})$.}
    \label{fig:BreakAdd}
\end{figure}

Recall from Section~\ref{sec:Relation_OpSpec_SymTFT} that there exist bi-local operators $\tr\psi_1(x)\psi_2(y)\in \mathcal{A}_\mathcal{F}(M_{d-1})$ or bi-local extended operators $\textrm{tr}\Psi_1(\Sigma_{d-2}) \Psi_2(\Sigma'_{d-2})$ when $\widetilde{\mathcal{L}} \neq \mathcal{L}$. When this is the case, we pick two disjoint regions $R_1$ and $R_2$ on the Cauchy surface $M_{d-1}$ and put one of the component operators of the bi-local operator, say $\psi_1(x)$ or $\Psi_1(\Sigma_{d-2})$, in $R_1$ while the other one $\psi_2(y)$ or $\Psi_2(\Sigma'_{d-2})$ in $R_2$ as illustrated in Figure~\ref{fig:BreakAdd}.
% \begin{figure}[!h]
%     \begin{equation}
%             \begin{gathered}
%         \begin{tikzpicture}[scale=0.7]
%             \draw[thick] (0,0)--(0,5)--(10,5)--(10,0)--(0,0);
%             \draw[thick,red] (0,2.5)--(5,2.5);
%             \draw[thick,blue] (5,2.5)--(10,2.5);
%             \filldraw[red] (2.5,2.5) circle (1.5pt); 
%             \filldraw[red] (7.5,2.5) circle (1.5pt); 
%             \node at (2.5,3) {$\mathcal{O}_1$};
%             \node at (7.5,3) {$\mathcal{O}_2$};  
%             \node at (2.5,1.8) {$R_1$};
%             \node at (7.5,1.8) {$R_2$};  
%             \node at (9,2.8) {$M_{d-1}$};
%             \node at (5,-0.5) {$M_{d}$};
%         \end{tikzpicture}
%     \end{gathered} \nonumber
%     \end{equation}
%     \caption{Putting a bi-local operator on $M_{d-1}$. Here the operator $\mathcal{O}$ can be either local operator $\psi(x)$ or extended operator $\Psi(\Sigma_{d-2})$.}
%     \label{fig:BreakAdd}
% \end{figure}
While $\psi_1(x),\Psi_1(\Sigma_{d-2})\notin \mathcal{A}_\mathcal{F}(R_1)$ and $\psi_2(y),\Psi_2(\Sigma'_{d-2})\notin \mathcal{A}_\mathcal{F}(R_2)$ as both are charged operators under $N$, we have
\begin{equation}
    \tr \psi_1(x)\psi_2(y) \in \mathcal{A}_\mathcal{F}(R_1\cup R_2)\,,\quad \textrm{tr}\Psi_1(\Sigma_{d-2}) \Psi_2(\Sigma'_{d-2})\in \mathcal{A}_\mathcal{F}(R_1\cup R_2)\,,
\end{equation}
due to~(\ref{eq:Inv_bi-local}) (or visually as Figure~\ref{Fig-bitwist-Operator} with $N = 1$). On the other hand, we have~\cite{Shao:2025mfj}
\begin{equation}
    \tr \psi_1(x)\psi_2(y) \notin (\mathcal{A}_\mathcal{F}(R_1)\cup \mathcal{A}_\mathcal{F}(R_2))''\,,\quad \textrm{tr}\Psi_1(\Sigma_{d-2}) \Psi_2(\Sigma'_{d-2})\notin (\mathcal{A}_\mathcal{F}(R_1)\cup \mathcal{A}_\mathcal{F}(R_2))''\,,
\end{equation}
the additivity of $\mathcal{A}_{\mathcal{F}}(M_{d-1})$ is violated. The reason is that we can construct a bi-twist operator with two endpoint operators lying in the blue region, and the symmetry generator connecting the two operators covers either $R_1$ or $R_2$. This operator is an element of the commutant $(\mathcal{A}_\mathcal{F}(R_1)\cup \mathcal{A}_\mathcal{F}(R_2))'$, and it does not commute with the above bi-local operators in general. By the discussion in Section~\ref{sec:Relation_OpSpec_SymTFT}, this means that addivity of $\mathcal{A}_\mathcal{F}(M_{d-1})$ is violated whenever $\widetilde{\mathcal{L}} \neq W_0 \oplus U_0$.

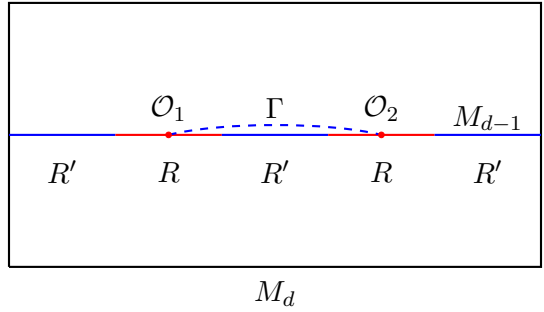
\begin{figure}[!h]
    \begin{equation}
            \begin{gathered}
        \begin{tikzpicture}[scale=0.7]
            \draw[thick] (0,0)--(0,5)--(10,5)--(10,0)--(0,0);
            \draw[thick,blue] (0,2.5)--(2,2.5);
            \draw[thick,red] (2,2.5)--(4,2.5);
            \draw[thick,blue] (4,2.5)--(6,2.5);
            \draw[thick,red] (6,2.5)--(8,2.5);
            \draw[thick,blue] (8,2.5)--(10,2.5);
            \draw[thick,blue,dashed] (3,2.5) .. controls (4,2.75) and (6,2.75) .. (7,2.5);
            \filldraw[red] (3,2.5) circle (1.5pt); 
            \filldraw[red] (7,2.5) circle (1.5pt); 
            \node at (3,3) {$\mathcal{O}_1$};
            \node at (7,3) {$\mathcal{O}_2$};  
            \node at (1,1.8) {$R'$};
            \node at (3,1.8) {$R$};
            \node at (5,1.8) {$R'$};
            \node at (7,1.8) {$R$};  
            \node at (9,1.8) {$R'$};
            \node at (5,3) {$\Gamma$};
            \node at (9,2.8) {$M_{d-1}$};
            \node at (5,-0.5) {$M_{d}$};
        \end{tikzpicture}
    \end{gathered} \nonumber
    \end{equation}
    \caption{Putting a bi-twist (extended) operator $\mathcal{O}_1 \Gamma \mathcal{O}_2$ on $M_{d-1}$. Here $\mathcal{O}_1 \Gamma \mathcal{O}_2$ can be either $\psi_1(x) \eta(\gamma_{x y}) \psi_2 (y)$ or $\Psi_1(\Sigma_{d-2}) N(\Sigma_{d-1}) \Psi_2(\Sigma'_{d-2})$. Note that though the tail $\Gamma$ connecting two operators is represented by the dashed curve slightly away from $M_{d-1}$ for clarity, we indeed have $\Gamma\subset M_{d-1}$.}
    \label{fig:NoBreakHaag}
\end{figure}

On the other hand, Haag duality can potentially be violated by the configuration in Figure~\ref{fig:NoBreakHaag} where $R\cup R' = M_{d-1}$ and $R\cap R' = \emptyset$.
Let us first focus on a bi-twist extended operator $\Psi_1(\Sigma_{d-2})N(\Sigma_{d-1})\Psi_2(\Sigma'_{d-2})$ plotted in Figure~\ref{fig:NoBreakHaag}, which is by definition not in $\mathcal{A}_\mathcal{F}(R)$ since its support is not entirely within $R$. Hence by~(\ref{eq:Haag}) Haag duality will be violated if $\Psi_1(\Sigma_{d-2})N(\Sigma_{d-1})\Psi_2(\Sigma'_{d-2})$ happens to commute with everything in $\mathcal{A}_\mathcal{F}(R')$, i.e. if it is in $\mathcal{A}_\mathcal{F}(R')'$.

Clearly, $\Psi_1(\Sigma_{d-2})N(\Sigma_{d-1})\Psi_2(\Sigma'_{d-2}) \notin \mathcal{A}_\mathcal{F}(R')'$ when $\mathcal{F}$ is invertible, i.e. when $\mathcal{F}$ is a group. To see this, we note that for $\mathcal{F}$ to be really a symmetry of $\mathcal{T}_\mathcal{F}$, there must exist at least one charged local operator that transforms non-trivially under $\mathcal{F}$. Thus one can form a bi-local operator $\tr \psi_{\mathbf{r}}(x) \psi_{\overline{\mathbf{r}}}(y)$ whose components $\psi_{\mathbf{r}}(x)$ and $\psi_{\overline{\mathbf{r}}}(y)$ both transform non-trivially under the action of $N$ but is itself in $\mathcal{A}_\mathcal{F}(M_{d-1})$. We then put $\psi_\mathbf{r}(x)$ in the middle $R'$ region and $\psi_{\overline{\mathbf{r}}}(y)$ in the leftmost $R'$ region in Figure~\ref{fig:NoBreakHaag}. Clearly, we have
\begin{equation}
    \tr \psi_\mathbf{r}(x)\psi_{\overline{\mathbf{r}}}(y) \in \mathcal{A}_{\mathcal{F}}(R')
\end{equation}
but since the charged local component $\psi_\mathbf{r}(x)$ does not commute with $N(\Sigma_{d-1})$, we have
\begin{equation}
    \Psi_1(\Sigma_{d-2})N(\Sigma_{d-1})\Psi_2(\Sigma'_{d-2}) \notin \mathcal{A}_{\mathcal{F}}(R')'\,.
\end{equation}
Therefore, when $\mathcal{F}$ is invertible, Haag duality of $\mathcal{A}_{\mathcal{F}}(M_{d-1})$ cannot be violated by any bi-twist extended operators. The argument also holds for bi-twist operator $\psi_1(x) \eta(\gamma_{xy}) \psi_2(y)$ where we can consider putting the bi-local extended operator $\textrm{tr}\Psi(\Sigma_{d-2}) \overline{\Psi}_{d-2}(\Sigma_{d-2})$ in different $R'$ region. 

It should now be clear that in order for Haag duality to potentially be violated in $\mathcal{A}_\mathcal{F}(M_{d-1})$, $\mathcal{F}$ has to be non-invertible. For example, to violate Haag duality by a twist operator $\Psi_1(\Sigma_{d-1}) N(\Sigma_{d-2})\Psi_2(\Sigma'_{d-1})$ we have to ensure that any uncharged bi-local operator $\tr \psi_1(x)\psi_2(y) \in \mathcal{A}_\mathcal{F}(M_{d-1})$ and bi-local extended operator $\tr \Psi_1(\Sigma_{d-2}) \Psi_2(\Sigma'_{d-2})$ have both $\psi_1(x),\Psi_1(\Sigma_{d-2})$ and $\psi_2(y),\Psi_2(\Sigma'_{d-2})$ invariant under $N(\Sigma)$, otherwise we would have $\Psi_1(\Sigma_{d-1}) N(\Sigma_{d-2})\Psi_2(\Sigma'_{d-1}) \notin \mathcal{A}_{\mathcal{F}}(R')'$ hence not violating Haag duality. Obviously, one possibility is that there are no non-trivial bi-local and bi-local extended operators in $\mathcal{A}_\mathcal{F}(M_{d-1})$ at all. This in turn means that any local and extended operator will become a twist operator upon the action of a non-invertible element of $\mathcal{F}$. Hence, it implies that $\widetilde{\mathcal{L}} = W_0 \oplus U_0$.

More generally, when $\widetilde{\mathcal{L}}\neq W_0 \oplus U_0$ and $\widetilde{\mathcal{L}} \neq \mathcal{L}$ to ensure $\mathcal{F}$ has non-invertible 0-form generators which can generate tails acting on the local (extended) operators. There are bi-local (extended) operators in $\mathcal{A}_\mathcal{F}(M_{d-1})$, which can be easily seen from the SymTFT construction in Section~\ref{sec:SvNA_SymTFT}. The question is then whether there exists an $N\in\mathcal{F}$ such that any local component of a bi-local (extended) operator in $\mathcal{A}_\mathcal{F}(M_{d-1})$ is invariant under the action of $N$. Recall that, by definition, for any $\tr \psi_1(x)\psi_2(y) \in \mathcal{A}_\mathcal{F}(M_{d-1})$, the component local operators $\psi_1(x)$ and $\psi_2(x)$ can never become a twist operator under any symmetry action in $\mathcal{F}$. In other words, a component local operator of an uncharged bi-local operator can at most undergo a linear transformation under any symmetry action in $\mathcal{F}$. Let us say it undergoes a linear transformation
\begin{equation}
    \psi_1(x) \rightarrow M(\Sigma_{d-1}) \psi_1(x) = \rho(M) \psi_1(x) M(\Sigma_{d-1})\,,
\end{equation}
under the action of certain 0-form symmetry $M\in\mathcal{F}$ and similarly for $\psi_2(y)$. Let us choose a non-invertible $M\in\mathcal{F}$ and consider the action of $\overline{M}M$ with $\overline{M}$ the orientation reversal of $M$, we have
\begin{equation}\label{Action-M-barM}
    \psi_1(x) \rightarrow \overline{M}(\Sigma_{d-1})M(\Sigma_{d-1}) \psi_1(x) =  \rho(M)\rho(\overline{M})\psi_1(x)\overline{M}(\Sigma_{d-1})M(\Sigma_{d-1})\,,
\end{equation}
where the fusion between $\overline{M}(\Sigma_{d-1})$ with $M(\Sigma_{d-1})$ generally gives
    \begin{equation}\label{Fusion-M-BarM}
        \overline{M}(\Sigma_{d-1}) M(\Sigma_{d-1}) = n_0  1 \oplus \left(\bigoplus_{N} N(\Sigma_{d-1})\right)\,,
    \end{equation}
where $N(\Sigma_{d-1})$ on the RHS can include genunie $(d-1)$-dimensional generator and also line operators $\eta(\gamma)$ condensing on $\Sigma_{d-1}$, as discussed in Section~\ref{sec:higher-dimensions}. Alternatively, we can first fuse $\overline{M}$ and $M$ and obtains
    \begin{equation}\label{M-Mbar-argument}
        \overline{M}(\Sigma)M(\Sigma) \psi_1(x) = \left(n_0  1 \oplus \left(\bigoplus_{N} N(\Sigma_{d-1})\right)\right)\psi_1(x) = \psi_1(x) n_0 1 \oplus \left(\bigoplus_{N} \rho(N)\psi_1(x)  N(\Sigma)\right)\,.
    \end{equation}
Compared to the RHS of \eqref{Action-M-barM}, we immediately have
    \begin{equation}
        \rho(M) \rho(\overline{M}) = \rho (N) = 1\,, \quad \forall N \in \overline{M} M\,,
    \end{equation}
which holds for any local operators $\psi_1(x)$ that cannot become a twist operator. Therefore, if there exists a bulk operator $U$ such that $N\in t(U)$ and $N \in \overline{M} M$ for some non-invertible symmetry $M$, we can construct the bi-twist operator $\Psi_1(\Sigma_{d-2})N(\Sigma_{d-1})\Psi_2(\Sigma'_{d-2})$ which commutes with $\tr \psi_1(x)\psi_2(y)$ for the configuration in Figure~\ref{fig:NoBreakHaag}. Notice that the above discussion is independent of the dimension of the operator and can be generalized to any charged objects $\mathcal{O}$ as long as $M$ is non-invertible such that \eqref{Fusion-M-BarM} holds. In particular, it also applies to the components of the bi-local extended operator $\tr \Psi_1(\Sigma_{d-2})\Psi_2(\Sigma'_{d-2}) \in \mathcal{A}_\mathcal{F}(M_{d-1})$.

In SymTFT, the set of operators in $\mathcal{L}$ gives natural candidates of the bi-twist operators in $\mathcal{A}_{\mathcal{F}}(M_{d-1})$ commuting with $\tr \psi_1(x)\psi_2(y)$ and $\tr \Psi_1(\Sigma_{d-2})\Psi_2(\Sigma'_{d-2})$. Since $\widetilde{\mathcal{L}}\neq \mathcal{L}$, we can always pick a bulk line operator $W'\in \mathcal{L}$ or $(d-1)$-dimensional extended operator $U'\in \mathcal{L}$ such that $W',U'\notin \widetilde{\mathcal{L}}$, and construct a symmetric bi-twist (extended) operator as illustrated in Section~\ref{sec:SvNA_SymTFT}. After projection to $\mathcal{T}_{\mathcal{F}}$, the former gives a bi-twist local operator connected by line operators $\eta'$, and the latter gives a bi-twist extended operator connected by $(d-1)$-dimensional generators $N'$. Since both $\eta'$ and $N'$ are tails that can end on some non-invertible generator $M$, they belong to $M \overline{M}$ and commute with all operators constructed by $\widetilde{\mathcal{L}}$ as we have just shown. 

As a result, we can rephrase the two statements at the beginning of this section in the language of SymTFT. Given a topological boundary $B_{\textrm{top}}$ defined by the set of endable operators (Lagrangian algebra) $\mathcal{L}$, we have
\begin{itemize}
    \item Additivity of $\mathcal{A}_{\mathcal{F}}(M_{d-1})$ is violated if $\widetilde{\mathcal{L}}\neq W_0 \bigoplus U_0$.
    \item Haag duality of $\mathcal{A}_{\mathcal{F}}(M_{d-1})$ is violated if $\widetilde{\mathcal{L}} \neq \mathcal{L}$.
    % the centralizer of $\widetilde{\mathcal{L}}_{\textrm{top}}$ is non-trivial
\end{itemize}
Here $\widetilde{\mathcal{L}}\subset \mathcal{L}$ is the subset of operators in $\mathcal{L}$ that cannot join to any non-trivial tail on $B_{\textrm{top}}$, and $W_0, U_0$ are respectively the identity line operator and the identity $(d-1)$-dimensional extended operator.

\subsection{Violation of Additivity and Haag Duality in 2d}
\label{sec:Add_Haag_Violation_2d}

Let us return to the 2D theory $\mathcal{T}_{\mathcal{F}}$ with 0-form symmetry $\mathcal{F}$, where the topological boundary $B_{\textrm{top}}$ is well classified by the Lagrangian algebra $\mathcal{L}$ \eqref{eq:Lagrangian_algebra} of the SymTFT. In this case, the statements in the previous section become
\begin{itemize}
    \item Additivity of $\mathcal{A}_{\mathcal{F}}(M_{1})$ is violated if $\widetilde{\mathcal{L}}\neq W_0$.
    \item Haag duality of $\mathcal{A}_{\mathcal{F}}(M_{1})$ is violated if $\widetilde{\mathcal{L}} \neq \mathcal{L}$.
    % the centralizer of $\widetilde{\mathcal{L}}_{\textrm{top}}$ is non-trivial
\end{itemize}

The violation of additivity is straightforward, since $\widetilde{\mathcal{L}} \neq 0$, we can construct symmetric bi-local operators using line operators in $\widetilde{\mathcal{L}}$ and thus the additivity is broken. On the other hand, suppose $\widetilde{\mathcal{L}}\neq \mathcal{L}$, there exists a bulk line operator $W \in \mathcal{L}$ but $W\notin \widetilde{\mathcal{L}}$, such that it can connect a tail $N$ on $B_{\textrm{top}}$. After projecting to $\mathcal{T}_{\mathcal{F}}$, that indicates the local operator $\psi$ corresponding to $W$ can be transformed to a twist operator $\psi$ attached by $N$ via some non-invertible symmetry $M$, and $N\in M \overline{M}$ since $N$ is attached to $M$ on the other end. The same argument in \eqref{M-Mbar-argument} implies $N$ must commute with any local operators in $\widetilde{\mathcal{L}}$, and thus the symmetric bi-twist operator $\psi t(W) \psi'$ constructed via $W$ can be used to break Haag duality.

We will provide another proof for the 2d theory to show the violation of Haag duality based on the following fact\footnote{We refer \cite{Barkeshli:2014cna} for a review of anyon systems.}. Suppose we have two line operators $W_{\alpha}, W_{\beta}$ in the bulk SymTFT linked with each other, and consider the quantity $S^*_{\alpha \beta} S_{00}/S_{0\alpha} S_{0\beta}$, where $S_{\alpha \beta}$ is the modular $S$-matrix in the bulk. If $S^*_{\alpha \beta} S_{00}/S_{0\alpha} S_{0\beta} = e^{i\phi(\alpha,\beta)}$ is a pure phase, then the braiding between $W_{\alpha}$ with $ W_{\beta}$ is abelian in the sense that
\begin{equation}\label{braiding}
    \begin{gathered}
        \begin{tikzpicture}
            \draw[thick] (0,-1)--(0,-0.5)--(0.2,-0.3);
            \draw[thick] (0.3,-0.2)--(0.5,0)--(0,0.5)--(0,1);
            \draw[thick] (0.5,-1)--(0.5,-0.5)--(0,0)--(0.2,0.2);
            \draw[thick] (0.3,0.3)--(0.5,0.5)--(0.5,1);
            \node at (-0.5,-1) {$W_{\alpha}$};
            \node at (1,-1) {$W_{\beta}$};
        \end{tikzpicture}
    \end{gathered}\quad = \quad \frac{S^*_{\alpha \beta}S_{00}}{S_{0\alpha} S_{0\beta}}\quad \times \quad 
    \begin{gathered}
        \begin{tikzpicture}
            \draw[thick] (0,-1)--(0,1);
            \draw[thick] (0.5,-1)--(0.5,1);
            \node at (-0.5,-1) {$W_{\alpha}$};
            \node at (1,-1) {$W_{\beta}$};
        \end{tikzpicture}     
    \end{gathered}\quad.
\end{equation}
In the following, we will show the linking between $W_{\alpha}\in \mathcal{\widetilde{L}}$ with $W_{\beta} \in \mathcal{L}$ is always trivial, namely
    \begin{equation}
        \frac{S^*_{\alpha \beta}S_{00}}{S_{0\alpha} S_{0\beta}}= 1\,, \quad \forall\  W_{\alpha} \in \widetilde{\mathcal{L}}\,, W_{\beta} \in \mathcal{L}\,.
    \end{equation}
It is a consequence of the fact that the line operators in the Lagrangian algebra are mutually local. If we further restrict $W_{\beta}\notin \widetilde{\mathcal{L}}$ and use $W_{\beta}$ to construct a symmetric bi-twist operator, then it should commute with any local operator $\psi$ constructed using $W_{\alpha} \in \widetilde{\mathcal{L}}$ by lifting the bi-twist operator to the bulk and moving it past through $W_{\alpha}$. 
  
To show $W_{\alpha}$ braid trivially with $W_{\beta}$ , we will make use of the following two properties of the Lagrangian algebra~\cite{Cong:2016ayp}. First, the quantum dimension of $\mathcal{L}$ is defined as
\begin{equation}
    \dim \mathcal{L} = \sum_{W_{\alpha}\in \mathcal{L}} n_{\alpha} d_{\alpha}\,,
\end{equation}
where $d_{\alpha}$ is the quantum dimension of $W_{\alpha}$. It turns out that $\dim \mathcal{L}$ is related to the total quantum dimension $D_{\mathcal{Z}(\mathcal{F})}$ of the SymTFT $\mathcal{Z}(\mathcal{F})$ as
\begin{equation}
    \dim \mathcal{L} = D_{\mathcal{Z}(\mathcal{F})}\,,
\end{equation}
where the quantum dimension $D_{\mathcal{Z}(\mathcal{F})}$ is defined by
\begin{equation}
    D_{\mathcal{Z}(\mathcal{F})} = \sqrt{\sum_{W_{\alpha}\in \mathcal{Z}(\mathcal{F})} d_{\alpha}^2}\,.
\end{equation}
Second, all the line operators $W_{\alpha}\in \mathcal{L}$ are bosonic. 

Recall that the modular $S$-matrix is given by
    \begin{equation}
        S_{\alpha \beta} = \frac{1}{D_{\mathcal{Z}(\mathcal{F})}} \sum_{\gamma} \mathcal{N}_{\alpha \beta}^{\gamma} \frac{\theta_{\gamma}}{\theta_{\alpha} \theta_{\beta}} d_{\gamma}\,,
    \end{equation}
where $\theta$ is the topological spin of $W$, which is the diagonal element of the modular $T$-matrix. If $W_{\alpha} \in \widetilde{\mathcal{L}}$, the fusion $W_{\alpha} \times \mathcal{L}$ should only contains elements inside $\mathcal{L}$, otherwise there exists a $W_{\gamma} \notin \mathcal{L}$ such that
    \begin{equation}
        \textrm{Hom}_{\mathcal{F}}(W_{\gamma},W_{\alpha}) = \textrm{Hom}_{\textrm{SymTFT}}(W_{\gamma},W_{\alpha}\times \mathcal{L})\neq 0 \,,
    \end{equation}
where we use \eqref{homorphism-coefficient}. That means $t(W_{\alpha})$ shares common elements with $t(W_{\gamma})$ on $B_{\textrm{top}}$. Since $W_{\gamma}$ must project to the non-trivial generator in $\mathcal{F}$ on the $B_{\textrm{top}}$, we have a contradiction. Therefore $W_\gamma\in W_{\alpha} \times W_{\mu}\in \mathcal{L}$ is also bosonic and we can set all topological spin $\theta_{\alpha},\theta_{\mu}$ and $\theta_{\gamma}$ equal to one and get
    \begin{equation}
        S_{\alpha \beta} = \frac{1}{D_{\mathcal{Z}(\mathcal{F})}} \sum_{\gamma} \mathcal{N}_{\alpha \beta}^{\gamma} d_{\gamma} = \frac{d_{\alpha} d_{\beta} }{D_{\mathcal{Z}(\mathcal{F})}}\,,
    \end{equation}
where we use the identity
    \begin{equation}
        d_{\alpha} d_{\mu} = \sum_{\gamma} \mathcal{N}_{\alpha \mu}^{\gamma} d_{\gamma}\,.
    \end{equation}
Therefore we have
    \begin{equation}
        \frac{S^*_{\alpha \beta}S_{00}}{S_{0 \alpha} S_{0\beta}} = \frac{d_{\alpha} d_{\beta} }{D_{\mathcal{Z}(\mathcal{F})}} \times \frac{S_{00}}{S_{0 \alpha}} \times \frac{S_{00}}{S_{0 \beta}}\times S_{00}^{-1}=1\,,
    \end{equation}
where we use $S_{0\alpha}/S_{00} = d_{\alpha}$ and $S_{00}=D^{-1}_{\mathcal{Z}(\mathcal{F})}$.

\section{Examples}\label{sec:examples}

In this section, we will study three types of 2D examples and illustrate the analysis of the violation of additivity and Haag duality from the SymTFT perspective. We will begin with the simplest abelian group $\mathbb{Z}_N$ and then move to general non-abelian groups and investigate $S_3$ as a concrete example. Then, we will consider the diagonal RCFT whose symmetries are generated by Verlinde lines and recover the criteria in \cite{Shao:2025mfj}.

\subsection{$\mathbb{Z}_N$ symmetry}

The first example is the $2D$ anomalous free $\mathbb{Z}_N$ theory $\mathcal{T}_{\mathbb{Z}_N}$. The SymTFT is the 3d BF-theory with level-$N$
    \begin{equation}
        S = \frac{N}{2\pi} \int B\wedge d A\,,
    \end{equation}
where we have a pair of $U(1)$-valued gauge field $A$ and $B$. The anyons are given by the gauge invariant Wilson line operators $W_{(\alpha,\beta)}[\Gamma]$ 
    \begin{equation}
        W_{(\alpha,\beta)} [\Gamma] = \exp \left(i \oint_{\Gamma} \alpha A + \beta B \right)\,,\quad (\alpha,\beta \in \mathbb{Z}_N)
    \end{equation}
with $\alpha,\beta = 0,1,\cdots,N-1$ and $\Gamma$ is a 1-cycle. The fusion rule is simply
\begin{equation}
    W_{(\alpha,\beta)} \times W_{(\alpha',\beta')} = W_{(\alpha+\alpha',\beta+\beta')}\,.
\end{equation}

The Lagrangian algebras can be found by finding the maximal commuting set of operators. We have two basic Lagrangian algebras, the Dirichlet one and the Neumann one
    \begin{equation}
        \mathcal{L}_{\textrm{Dir}} = \bigoplus_{\alpha\in \mathbb{Z}_N} W_{(\alpha,0)}\,,\quad \mathcal{L}_{\textrm{Neu}} =  \bigoplus_{\beta\in \mathbb{Z}_N} W_{(0,\beta)}\,.
    \end{equation}
If $N$ is not prime, for any positive integers $P,Q$ satisfying $N=PQ$ and $\gcd(P,Q)\neq 1$, one has others Lagrangian algebras
    \begin{equation}
        \mathcal{L}_{P,Q} = \bigoplus_{\alpha\in \mathbb{Z}_Q,\beta\in \mathbb{Z}_P} W_{(P\alpha,Q\beta)}\,,
    \end{equation}
and the Dirichlet (Neumann) algebra is the special case when $P=1$ ($Q=1$).

Begin with the topological boundary $B_{\textrm{Dir}}$ where the anyons $W_{(\alpha,0)}$ can end on that, it supports a $\mathbb{Z}_N$ symmetry generated by $W_{(0,1)}$. The Neumann boundary $B_{\textrm{Neu}}$ is obtained by gauging the $\mathbb{Z}_N$ symmetry generated by $W_{(0,1)}$ and the corresponding Lagrangian algebra becomes $\mathcal{L}_{\textrm{Neu}}$. The role between $W_{(\alpha,0)}$ and $W_{(0,\beta)}$ are exchanged on $B_{\textrm{Neu}}$ such that $W_{(0,\alpha)}$ can end on the boundary while $W_{(1,0)}$ serves as symmetry generator of the dual $\mathbb{Z}_N$ symmetry. Moreover, if we only gauge a subgroup of $\mathbb{Z}_P \in \mathbb{Z}_N$ we will obtain the topological boundary $B_{(P,Q)}$ whose Lagrangian algebra is $\mathcal{L}_{(P,Q)}$. The symmetry supported on $B_{(P,Q)}$ is $\mathbb{Z}_P \times \mathbb{Z}_Q$ with a mixed anomaly between the two factors, generated separately by $W_{(1,0)}$ and $W_{(0,1)}$.

For all of them, one finds
    \begin{equation}
        \widetilde{\mathcal{L}}_{\textrm{Dir}}=\mathcal{L}_{\textrm{Dir}}\,,\quad \widetilde{\mathcal{L}}_{\textrm{Neu}}=\mathcal{L}_{\textrm{Neu}}\,,\quad \widetilde{\mathcal{L}}_{P,Q}=\mathcal{L}_{P,Q}\,.
    \end{equation}
Therefore, the additivity is violated, and the Haag duality is preserved. It matches our result for invertible $\mathcal{F}$, i.e. when $\mathcal{F}$ is a group.

\subsection{$S_3$ symmetry}

For a general discrete group $G$, the elements of its SymTFT $\mathcal{Z}(G)$ are described by the pair $\alpha = (A,r)$ where $A$ labels the conjugacy class $Cl(G)$ of $G$ and $r$ labels the irreducible representation under the centralizer group of $Cl(G)$. 

Let us consider the non-anomalous order-3 cyclic group $S_3 = \{e,a,a^2,b,ab,a^2 b \}$. $a$ and $b$ are the generators of $\mathbb{Z}_3$ and $\mathbb{Z}_2$ satisfying $a^3=b^2=1$ and the constraint $ba = a^2 b$. There are three conjugacy classes represented by $[e],[a],[b]$
    \begin{equation}
        [e] = \{ e \}\,,\quad [a] = \{ a , a^2 \}\,,\quad [b] = \{ b , ab, a^2 b\}\,,
    \end{equation}
and the centralizer group for $e,a,b$ are separately
    \begin{equation}
        C(e) = S_3\,,\quad C(a) = \mathbb{Z}_3\,, \quad C(b) = \mathbb{Z}_2 \,,
    \end{equation}
where $\mathbb{Z}_3$ and $\mathbb{Z}_2$ are generated by $a$ and $b$ separately. Recall that the irreducible representations for $\mathbb{Z}_N$ are characterized by an $\mathbb{Z}_N$-integer $k$ with
    \begin{equation}
        r_k (g) = e^{\frac{2\pi i}{N} k},\quad \textrm{($g$ is the generator of $\mathbb{Z}_N$)}
    \end{equation}
and the irreducible representations for $S_3$ contain the trivial representation $r_+$, sign representation $r_-$ and the 2-dimensional standard representation $r_E$. There are in total 8 line operators labeled by
    \begin{equation}
        ([e],r_+),([e],r_-),([e],r_E),([b],r_0),([b],r_1),([a],r_0),([a],r_1),([a],r_2)
    \end{equation}
and we shall label them as $W_0,\cdots,W_7$ in the following. The fusion rule is summarized in Table~\ref{tab:my_label}.
    \begin{table}[!h]
    \footnotesize
    \hspace{-1.5cm}
    \begin{tabular}{|c|c|c|c|c|c|c|c|c|}
    \hline
    $\otimes$& $W_0$ & $W_1$ &$W_2$&$W_3$&$W_4$&$W_5$&$W_6$&$W_7$\\
    \hline
    $W_0$& $W_0$ & $W_1$ &$W_2$&$W_3$&$W_4$&$W_5$&$W_6$&$W_7$\\
    \hline
    $W_1$&$W_1$&$W_0$&$W_2$&$W_4$&$W_3$&$W_5$&$W_6$&$W_7$\\
    \hline
    $W_2$&$W_2$&$W_2$&$W_0\oplus W_1\oplus W_2$&$W_3\oplus W_4$&$W_3\oplus W_4$&$W_6\oplus W_7$&$W_5\oplus W_7$&$W_5\oplus W_6$\\
    \hline
    $W_3$&$W_3$&$W_4$&$W_3\oplus W_4$&$\makecell{W_0\oplus W_2 \oplus W_5\\ \oplus W_6 \oplus W_7}$ & $\makecell{W_1\oplus W_2 \oplus W_5\\ \oplus W_6 \oplus W_7}$&$W_3\oplus W_4$&$W_3\oplus W_4$&$W_3\oplus W_4$\\
    \hline
    $W_4$&$W_4$&$W_3$&$W_3\oplus W_4$&$\makecell{W_1\oplus W_2 \oplus W_5\\ \oplus W_6 \oplus W_7}$&$\makecell{W_0\oplus W_2 \oplus W_5\\ \oplus W_6 \oplus W_7}$&$W_3\oplus W_4$&$W_3\oplus W_4$&$W_3\oplus W_4$\\
    \hline
    $W_5$&$W_5$&$W_5$&$W_6\oplus W_7$&$W_3\oplus W_4$&$W_3\oplus W_4$&$W_0\oplus W_1 \oplus W_5$&$W_7 \oplus W_2$&$W_6\oplus W_2$\\
    \hline
    $W_6$&$W_6$&$W_6$&$W_5\oplus W_7$&$W_3\oplus W_4$&$W_3\oplus W_4$&$W_7\oplus W_2$&$W_0 \oplus W_1 \oplus W_6$&$W_5\oplus W_2$\\
    \hline
    $W_7$&$W_7$&$W_7$&$W_5\oplus W_6$&$W_3\oplus W_4$&$W_3\oplus W_4$&$W_6\oplus W_2$&$W_5 \oplus W_2$&$W_0\oplus W_1\oplus W_7$\\
    \hline
    \end{tabular}
    \caption{The fusion table of $\mathcal{Z}(S_3)$.}
    \label{tab:my_label}
\end{table}

There are four Lagrangian algebras given by\cite{Cong:2016ayp}
    \begin{gather}
        \begin{split}
        \mathcal{L}_1 &= W_0 \oplus W_1 \oplus 2 W_2\,,\quad
        \mathcal{L}_2 = W_0 \oplus W_3 \oplus W_5\,,\\
        \mathcal{L}_3 &= W_0 \oplus W_1 \oplus 2W_5\,,\quad
        \mathcal{L}_4 = W_0 \oplus W_2 \oplus W_3\,.
        \end{split}        
    \end{gather}
For each Lagrangian algebra $\mathcal{L}_i$, we have the corresponding topological boundary $B_i$, and they are related as follows. Begin with $\mathcal{L}_1$ where $W_0, W_1,W_2$ can end on the topological boundary $B_{1}$, it supports a $S_3$ symmetry whose generators will be identified later. If we gauge $S_3$ on $B_{1}$, we will obtain $B_2$ characterized by $\mathcal{L}_2$, and the underlying symmetry is $\textrm{Rep}(S_3)$. Alternatively, we can also gauge $\mathbb{Z}_3\in S_3$ or $\mathbb{Z}_2\in S_3$, and we will get separately $B_3$ and $B_4$ whose Lagrangian algebras are $\mathcal{L}_3$ and $\mathcal{L}_4$. The symmetries supported on $B_3$ and $B_4$ are still $S_3$ and $\textrm{Rep}(S_3)$.

In order to determine the tails of each line operator on the topological boundary, we need to consider the homorphism introduced in \eqref{homorphism-coefficient} and thus consider the fusion $W_{\alpha} \times \mathcal{L}_i$. It is convenient to label each line operator $W_{\alpha}$ using 8-dimensional orthogonal basis such that
    \begin{equation}
        |W_\alpha\rangle\quad  \equiv \underbrace{|0,\cdots 0,1,0,\cdots\rangle}_\text{The $(\alpha+1)$-th component is $1$}
    \end{equation}
and one can write
    \begin{equation}
        \begin{split}
        |\mathcal{L}_1\rangle =& |1,1,2,0,0,0,0,0\rangle\,,\quad
        |\mathcal{L}_2\rangle = |1,0,0,1,0,1,0,0\rangle\,,\\
        |\mathcal{L}_3\rangle =& |1,1,0,0,0,2,0,0\rangle\,,\quad
        |\mathcal{L}_4\rangle = |1,0,1,1,0,0,0,0\rangle\,.
        \end{split}
    \end{equation}
Then the dimension of homomorphism in $\mathcal{Z}(S_3)$ is $\dim \textrm{Hom}_{\mathcal{Z}(S_3)}(W_{\alpha},W_{\beta}) = \langle W_{\alpha} | W_{\beta}\rangle$, and the dimension of homomorphism in $\mathcal{F}(B_i)$ is
    \begin{equation}
        \dim \textrm{Hom}_{\mathcal{F}(B_i)}(t(W_{\alpha}),t(W_{\beta})) = \langle W_{\alpha} | W_{\beta} \times \mathcal{L}_i\rangle\,.
    \end{equation}
We will mainly focus on $\mathcal{L}_1$ and $\mathcal{L}_2$, and the discussion of $\mathcal{L}_3$ and $\mathcal{L}_4$ are totally parallel to $\mathcal{L}_1$ and $\mathcal{L}_2$.

\subsubsection*{Topological boundary $B_1$ : $S_3$-Symmetry}

One can work out the fusion between each line operator $W_{\alpha}$ with $\mathcal{L}_1$ as summarized in Table \ref{S3-A1}. 
    \begin{table}[!h]
        \centering
        \begin{tabular}{|c|c|}
        \hline
            $W_{\alpha}$  & $|W_{\alpha} \times \mathcal{L}_{1}\rangle$ \\
            \hline
            $W_0$ &$|1,1,2,0,0,0,0,0\rangle$\\
            \hline
            $W_1$ &$|1,1,2,0,0,0,0,0\rangle$\\
            \hline
            $W_2$ &$|2,2,4,0,0,0,0,0\rangle$\\
            \hline
            $W_3$ & $|0,0,0,3,3,0,0,0\rangle$\\
            \hline
            $W_4$ & $|0,0,0,3,3,0,0,0\rangle$\\
            \hline
            $W_5$ & $|0,0,0,0,0,2,2,2\rangle$\\
            \hline
            $W_6$ & $|0,0,0,0,0,2,2,2\rangle$\\
            \hline
            $W_7$ & $|0,0,0,0,0,2,2,2\rangle$\\
            \hline
        \end{tabular}
        \caption{The fusion between line operators with $\mathcal{L}_1$}
        \label{S3-A1}
    \end{table}
One can see $2W_0,2W_1,W_2$ are all the same after fusing $\mathcal{L}_1$, and one can write
    \begin{equation}
        \textrm{Hom}_{\mathcal{F}(B_1)} (t(W_{\alpha}) ,t(2W_0)) =\textrm{Hom}_{\mathcal{F}(B_1)} (t(W_{\alpha}) ,t(2W_1)) =\textrm{Hom}_{\mathcal{F}(B_1)} (t(W_{\alpha}) ,t(W_2))\,,
    \end{equation}
for any $W_{\alpha} \in \mathcal{Z}(S_3)$. That indicates $2W_0, 2W_1, W_2$ project to the same elements on the topological boundary. Since $t(W_0)$ is the identity element in $\mathcal{F}(\mathcal{L}_1)$, we can conclude
    \begin{equation}
        t(W_0) = t(W_1) = 1\,,\quad t(W_2) = 1\oplus 1\,.
    \end{equation}
Similarly, we also have the following identifications
    \begin{equation}
         t(W_3) = t(W_4)\,,\quad t(W_5) = t(W_6) = t(W_7)\,,
    \end{equation}
on the topological boundary.
We also need to determine whether they are simple or not at the boundary. Using \eqref{homorphism-coefficient} it is easy to find
    \begin{equation}
        \dim \textrm{Hom}_{\mathcal{F}(\mathcal{L}_1)}(t(W_{3}),t(W_{3})) =  3\,,\quad
        \dim \textrm{Hom}_{\mathcal{F}(\mathcal{L}_1)}(t(W_{5}),t(W_{5})) = 2\,,
    \end{equation}
which indicates $W_3$ is composed of three simple elements and $W_5$ is composed of two simple elements on the boundary, and we can write
    \begin{equation}
        t(W_3) = \beta_1 \oplus \beta_2 \oplus \beta_3\,,\quad t(W_5) = \alpha_1 \oplus \alpha_2\,.
    \end{equation}
Moreover, the fusion rules between $W_3$ and $W_5$ in Table.~\ref{tab:my_label} implies the following fusion rules on $B_1$
\begin{equation}
        \begin{gathered}
        (\alpha_1 \oplus \alpha_2) \times (\alpha_1 \oplus \alpha_2) = 1 \oplus 1 \oplus \alpha_1 \oplus \alpha_2\,,\\
        (\beta_1 \oplus \beta_2 \oplus \beta_3)^2 = 3 \oplus \alpha_1 \oplus \alpha_2\,,\\ (\beta_1 \oplus \beta_2 \oplus \beta_3) \times (\alpha_1\oplus\alpha_2) = 2\beta_1 \oplus 2\beta_2 \oplus 2\beta_3\,.
        \end{gathered}
\end{equation}
The first one is satisfied if we think of $\alpha_1,\alpha_2$ generating a $\mathbb{Z}_3$ symmetry and identify $\alpha_1 = a,\alpha_2 = a^2$. If we further identify
    \begin{equation}
        \beta_1 = b,\quad \beta_2 = ab,\quad \beta_3 = a^2b,
    \end{equation}
where $b$ is the $\mathbb{Z}_2$ generator satisfying satisfying $ba=a^2b$, then they solve the second and third fusion rules. Therefore, we can deduce the symmetry $\mathcal{F}(\mathcal{L}_1)$ on $B_1$ is $S_3$ and $W_0,W_3, W_5$ are conjugacy classes $[e],[a],[b]$ when restricting to the boundary. 

Since the fusions between $W_0,W_1,W_2$ with $\mathcal{L}_1$ are all inside $\mathcal{L}_1$, we have $\widetilde{\mathcal{L}}_1 = \mathcal{L}_1$ so that the additivity is violated but the Haag duality is preserved.

\subsubsection*{Topological boundary $B_2$ : $\textrm{Rep}(S_3)$}

One can work out the fusion between each line operator $W_{\alpha}$ with $\mathcal{L}_2$ as summarized in Table~\ref{S3-A4}. 
    \begin{table}[!h]
        \centering
        \begin{tabular}{|c|c|}
        \hline
            $W_{\alpha}$ & $|W_{\alpha}\times \mathcal{L}_2\rangle$ \\
            \hline
            $W_0$ &$|1, 0, 0, 1, 0, 1, 0, 0\rangle$\\
            \hline
            $W_1$ &$|0, 1, 0, 0, 1, 1, 0, 0\rangle$\\
            \hline
            $W_2$ &$|0, 0, 1, 1, 1, 0, 1, 1\rangle$\\
            \hline
            $W_3$  & $|1, 0, 1, 2, 1, 1, 1, 1\rangle$\\
            \hline
            $W_4$  & $|0, 1, 1, 1, 2, 1, 1, 1\rangle$\\
            \hline
            $W_5$  & $|1, 1, 0, 1, 1, 2, 0, 0\rangle$\\
            \hline
            $W_6$ & $|0, 0, 1, 1, 1, 0, 1, 1\rangle$\\
            \hline
            $W_7$  & $|0, 0, 1, 1, 1, 0, 1, 1\rangle$\\
            \hline
        \end{tabular}
        \caption{The fusion between line operators with $\mathcal{L}_2$}
        \label{S3-A4}
    \end{table}
And we should identify the following line operators on the topological boundary
    \begin{equation}
    \begin{gathered}
        t(W_2) = t(W_6) = t(W_7)\,,\quad t(W_5) = t(W_0) + t(W_1)\,,\\
        t(W_3) = t(W_0) + t(W_2)\,,\quad t(W_4) = t(W_1) + t(W_2)\,.      
    \end{gathered}
    \end{equation}
We can pick three generators $t(W_0),t(W_1),t(W_2)$. Here $t(W_0)=1$ is the identity, and from \eqref{homorphism-coefficient} we can read the multiplicity
    \begin{equation}
        \dim \textrm{Hom}_{\mathcal{F}(\mathcal{L}_2)}(t(W_1),t(W_1))  = 1\,,\quad
        \dim \textrm{Hom}_{\mathcal{F}(\mathcal{L}_2)}(t(W_2),t(W_2)) = 1\,,
    \end{equation}
therefore all of the three line operators are simple on the boundary. Let us denote $t(W_1) = \eta$ and $t(W_2) =E$, the fusion rules between $W_1$ and $W_2$ in Table.~\ref{tab:my_label} further implies
    \begin{equation}
        \eta\times \eta = 1\,,\quad 
        \eta\times E=E\,,\quad
        E\times E=1 \oplus  \eta \oplus E\,,
    \end{equation}
which are exactly the fusion rules of Rep($S_3$): $1$ is the trivial representation $r_+$, $\eta$ and $E$ are the sign representation $r_-$ and the 2-dimensional representation $r_E$.

\begin{figure}[!h]
    \begin{equation}
            \begin{gathered}
        \begin{tikzpicture}
            \draw[thick,blue] (1,-1)--(1,1);
            \filldraw[red] (0,0) circle (1.5pt); 
            \node at (1,-1.5) {$E$};
            \node at (0,-0.5) {$\psi_{[b],r_0}$};
            \node at (2.5,0) {$\Rightarrow$};
        \end{tikzpicture} \qquad
        \begin{tikzpicture}
            \draw[thick,blue] (-1,-1)--(-1,1);
            \draw[thick,blue] (-1,0)--(0,0);
            \filldraw[red] (0,0) circle (1.5pt); 
            \node at (-1,-1.5) {$E$};
            \node at (-0.5,0.25) {$E$};
            \node at (0,-0.5) {$\psi'_{[b],r_0}$};
        \end{tikzpicture}\\
        \begin{tikzpicture}
            \draw[thick,blue] (1,-1)--(1,1);
            \filldraw[red] (0,0) circle (1.5pt); 
            \node at (1,-1.5) {$E$};
            \node at (0,-0.5) {$\psi_{[a],r_0}$};
            \node at (2.5,0) {$\Rightarrow$};
        \end{tikzpicture} \qquad
        \begin{tikzpicture}
            \draw[thick,blue] (-1,-1)--(-1,1);
            \filldraw[red] (0,0) circle (1.5pt); 
            \node at (-2,0) {$-\frac{1}{2}$};
            \node at (-1,-1.5) {$E$};
            \node at (0,-0.5) {$\psi'_{[a],r_0}$};
            \node at (1,0) {$+$};
            \node at (2,0) {$\left(\omega +\frac{1}{2}\right)$};
            \draw[thick,blue] (3,-1)--(3,1);
            \draw[thick,dashed,blue] (3,0)--(4,0);
            \filldraw[red] (4,0) circle (1.5pt); 
            \node at (4,-0.5) {$\psi'_{[a],r_0}$};
            \node at (3,-1.5) {$E$};
            \node at (3.5,0.25) {$\eta$};
        \end{tikzpicture}
    \end{gathered}\nonumber 
    \end{equation}
    \caption{The local operators can transform to twist operators under the $E$-transformation in $\textrm{Rep}(S_3)$. Here $\omega = e^{2\pi i/3}$ is the 3rd-root of unity.}
    \label{twist-operator-RepS3}
\end{figure}
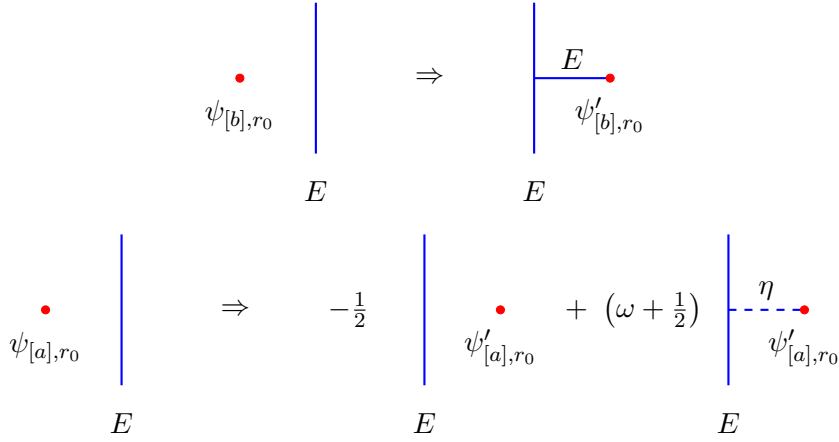

Since we have $t(W_3)=1 \oplus E$ and $t(W_5) = 1 \oplus \eta$, therefore $\widetilde{\mathcal{L}}_{2} = W_0$. One may conclude that the additivity is preserved, but the Haag duality is violated. Indeed, if we denote the local operators represented by $W_3$ and $W_5$ as $\psi_{[b],r_0}$ and $\psi_{[a],r_0}$, they will transform under the non-invertible $E$-symmetry according to Figure~\ref{twist-operator-RepS3} \cite{Bhardwaj:2023idu}.
Since both $\psi_{[b],r_0}$ and $\psi_{[a],r_0}$ form multiplets with twist operators, one cannot build a symmetric bi-local operator to violate the additivity. On the other hand, we can easily construct a symmetric bi-twist operator using bulk lines so that the Haag duality is violated. Notice that this is an example where we have an invertible symmetry $\eta \in \textrm{Rep}(S_3)$ but the additivity is not violated.

\subsection{Diagonal RCFT}
Finally, we will revisit the diagonal RCFT whose symmetry $\mathcal{F}$ is generated by the Verlinde lines, and show that the two conditions are the same as those in \cite{Shao:2025mfj}. The local primary operators of the diagonal RCFT are denoted as $\phi_{\alpha,\alpha^*}(z,\bar{z})$ with conformal weight $(h_{\alpha},h_{\alpha})$. They are one-to-one correspondent to the Verlinde lines $L_{\alpha}$ in the diagonal RCFT with the topological spin $\theta_{\alpha} = \exp\left(2\pi ih_{\alpha}\right)$.

The Verlinde lines generate a non-invertible symmetry $\mathcal{F}$, and the SymTFT $\mathcal{Z}(\mathcal{F})$ of $\mathcal{F}$ is the quantum double defined by
    \begin{equation}
        \mathcal{Z}(\mathcal{F}) = \mathcal{F}\boxtimes \bar{\mathcal{F}}\,,
    \end{equation}
where the bulk anyons $W_{(\alpha,\beta)}$ are labeled by a pair $(\alpha,\beta)$ as
    \begin{equation}
        W_{(\alpha,\beta)} \equiv L_{\alpha} \boxtimes L_{\bar{\beta}}\,.
    \end{equation}
In the SymTFT picture, the local primary operators $\phi_{\alpha,\alpha^*}(z,\bar{z})$ are considered as the anyon $W_{(\alpha,\alpha)}$ emitting from an operator in the physical boundary and ending on the topological boundary $B_{\textrm{top}}$. Therefore, the Lagrangian algebra for a diagonal RCFT is
    \begin{equation}
        \mathcal{L}_{\textrm{Diag}} = \bigoplus_{\alpha} W_{(\alpha,\alpha)} = \bigoplus_{\alpha} L_{\alpha} \boxtimes L_{\bar{\alpha}}\,.
    \end{equation}
On the other hand, the non-diagonal anyons $W_{(\alpha,\beta)}$ cannot end on $B_{\textrm{top}}$ and will connect to non-trivial Verlinde lines on the topological boudary. They will lead to the twist primaries $\phi_{\alpha,\beta^*}(z,\bar{z})$ with $\alpha\neq \beta$, which is attached to some Verlinde line $L_{\gamma}$ in the diagonal RCFT.

The subalgebra $\widetilde{\mathcal{L}}_{\textrm{Diag}}$ is given by the subset of diagonal anyon $W_{(\alpha,\alpha)}$ satisfying
    \begin{equation}
        W_{(\alpha,\alpha)} \times \mathcal{L}_{\textrm{Diag}} \in \mathcal{L}_{\textrm{Diag}}\,, 
    \end{equation}
so that it cannot transit to any non-trivial Verlinde lines on the boundary according to \eqref{homorphism-coefficient}.
In particular, the fusion between $W_{\alpha,\alpha}$ with $W_{\bar{\alpha},\bar{\alpha}}\in \mathcal{L}_{\textrm{Diag}}$ implies $L_{\alpha}$ is invertible
    \begin{equation}
        L_{\alpha}\times L_{\bar{\alpha}} = 1\,,
    \end{equation}
otherwise there will be non-diagonal anyons like $W_{(0,\gamma)}$ and $W_{(\gamma,0)}$ generated in the fusion $ W_{(\alpha,\alpha)} \times W_{(\bar{\alpha},\bar{\alpha})}$. One can also prove $W_{(\alpha,\alpha)} \in \widetilde{L}_{\textrm{Diag}}$ is equivalent to $L_{\alpha}\times L_{\bar{\alpha}} = 1$. Suppose the latter is true, which means the fusion coefficient satisfies
    \begin{equation}
        N_{\alpha \bar{\alpha}}^{\beta} = \left\{  \begin{array}{l}
            1\,, \quad (\beta=0)\\
            0\,, \quad (\beta \neq 0)
        \end{array}   
        \right.
    \end{equation}
Notice that the fusion coefficient satisfies the identity
    \begin{equation}
         \sum_f N_{af}^d N_{bc}^f=\sum_e N_{ab}^e N_{ec}^d\,.
    \end{equation}
Setting $a=\bar{\alpha},b=\alpha$ one has
    \begin{equation}
        \sum_{f} N_{\bar{\alpha} f}^d N_{\alpha c}^f = \delta_c^d\,,
    \end{equation}
where we used $N_{0c}^d = \delta_c^d$. Then sum over $d$ we have
    \begin{equation}
        \sum_{f}\left(\sum_d N_{\bar{\alpha} f}^d\right) N_{\alpha c}^f = 1\,,
    \end{equation}
which holds for any $c$. The point is one cannot have $L_{\bar{\alpha}} \times L_{f} = 0$ since that would imply $L_{f} = 0$ by multiplying $L_{\alpha}$ on both side, therefore one must have $\sum_d N_{\bar{\alpha} f}^d \in \mathbb{Z}_{\geq1}$ for any $f$. Since the fusion coefficients $N_{\alpha c}^f$ are also non-negative integers, that implies there exist a $\gamma(c)$ depending on $c$ such that
\begin{equation}
     \sum_d N_{\bar{\alpha} \gamma(c)}^d = 1\,,\quad N_{\alpha c}^f = \left\{\begin{array}{l}
        1\,,\quad f = \gamma(c)\,,\\
        0\,,\quad f\neq \gamma(c)\,.
    \end{array} \right.
\end{equation}
Thus we arrive at
    \begin{equation}
        L_{\alpha} \times L_{c} = L_{\gamma(c)}\,,
    \end{equation}
which means the fusion between $L_{\alpha}$ with any other Verlinde line $L_{c}$ will map that into a single Verlinde line $L_{\gamma(c)}$, and one has
    \begin{equation}
        W_{(\alpha,\alpha)} \times W_{(\beta,\beta)} = W_{(\beta(c),\beta(c))} \in \mathcal{L}_{\textrm{Diag}}\, \quad \Rightarrow \, \quad W_{(\alpha,\alpha)} \times \mathcal{L}_{\textrm{Diag}} \in \mathcal{L}_{\textrm{Diag}}\,.
    \end{equation}

Therefore $\widetilde{\mathcal{L}}_{\textrm{Diag}}$ contains all the diagonal anyons $W_{(\alpha,\alpha)}=L_{\alpha}\boxtimes L_{\bar{\alpha}}$ with invertible Verlinde lines $L_{\alpha}$
    \begin{equation}
        \widetilde{\mathcal{L}}_{\textrm{Diag}} = \bigoplus_{\textrm{invertible }L_{\alpha}} L_{\alpha}\boxtimes L_{\bar{\alpha}}\,.
    \end{equation}
Let us denote $C_{\textrm{ab}}\in \mathcal{F}$ as the set generated by the invertible Verlinde lines, then the addivitiy is violated if $\widetilde{\mathcal{L}}_{\textrm{Diag}}\neq L_0 \boxtimes L_0$, or equivalently $C_{\textrm{ab}}$ contains Verlinde lines other than $L_0$. That indicates $\mathcal{F}$ contains non-trivial invertible symmetries. On the other hand, the Haag duality is violated when $\widetilde{\mathcal{L}}_{\textrm{Diag}}\neq \mathcal{L}_{\textrm{Diag}}$, or equivalently when $C_{\textrm{ab}}\subsetneq \mathcal{F}$. Which means $\mathcal{F}$ contains non-invertible symmetry generators.

\section{Conclusion and Discussion}

In this work, we generalize the result of~\cite{Shao:2025mfj} by exploiting the power of SymTFT in the analysis of the global symmetries of the QFT. The many edge cases when attempting the construction of uncharged non-local operators purely by analyzing the symmetry algebra of QFT as exemplified in Section~\ref{sec:SvNA_OpSpec} are once-and-for-all resolved by analyzing the Lagrangian algebra of the corresponding SymTFT in a simple and unified way as shown in Section~\ref{sec:SvNA_SymTFT}. In Section~\ref{sec:Add_Haag_Violation} we summarize and prove the conditions for the violation of additivity and Haag duality of the von Neumann algebra as certain constraints on the Lagrangian algebra, which offers an immediate generalization to any dimension. We then provide concrete examples in the SymTFT language to cross-check our statements in Section~\ref{sec:examples}.

The key observation is that the violation of additivity or Haag duality of the von Neumann algebra can be detected when restricting the von Neumann algebra to its uncharged sector and including in it certain types of non-local operators. The operators we consider are either bi-local or bi-twist operators, the former is an uncharged pair of charged local operators (or extended operators each of which projects to a local operator when projecting its support down to a point) while the latter is an uncharged combination of two charged operators connected by an operator labeled by an element of $\mathcal{F}$ supported on a manifold whose two boundaries support the aforementioned two charged operators. The neutralness of such non-local operators can then be determined conveniently by analyzing the corresponding operator in the SymTFT as graphically displayed in Figure~\ref{Fig-bitwist-Operator} and~\ref{Fig-patch-Operator}. We define the notion of the tail of a bulk operator to facilitate the analysis and the set $\widetilde{\mathcal{L}}\subset \mathcal{L}$ of the bulk operators that cannot join to non-trivial tails on the topological boundary. Using the segmentation of the Cauchy surface proposed in~\cite{Shao:2025mfj}, we summarize and prove the condition for the violation of additivity as $\widetilde{\mathcal{L}}\neq W_0 \oplus U_0$ and for the violation of Haag duality as $\widetilde{\mathcal{L}}\neq \mathcal{L}$.

An interesting future direction would be to generalize the current framework to spacetime symmetries. Recently, the author of~\cite{Pace:2025hpb} has made progress on incorporating spacetime symmetry in SymTFT, resulting in symmetry-enriched topological (SET) order. The challenge lies in the construction of the non-local operators in such SymTFT with spacetime symmetry data and we would like to attempt this in future works.

% \appendix
% \section{Some title}
% Please always give a title also for the appendices.

\acknowledgments

We would like to thank Ziming Ji, Yi-Nan Wang, Zhenbin Yang and Yi Zhang for helpful discussions. JT would like to thank the Peng Huanwu Center for Fundamental Theory for hosting ``Some formal aspects of Field Theories and Holography'' where the work was first presented and the excellent questions raised by Wei Li, Yi-Nan Wang and Zhenbin Yang. QJ is supported by National Research Foundation of Korea (NRF) Grant No. RS-2024-00405629 and Jang Young-Sil Fellow Program at the Korea Advanced Institute of Science and Technology. JT is supported by National Natural Science Foundation of China under Grant No. 12405085 and by the Natural Science Foundation of Shanghai (Grant No. 24ZR1419300).

% \paragraph{Note added.} This is also a good position for notes added after the paper has been written.

% Bibliography

%% [A] Recommended: using JHEP.bst file
\bibliographystyle{JHEP}
\bibliography{refs.bib}

%% or
%% [B] Manual formatting (see below)
%% (i) We suggest to always provide author, title and journal data or doi:
%% in short all the informations that clearly identify a document.
%% (ii) please avoid comments such as "For a review'', "For some examples",
%% "and references therein" or move them in the text. In general, please leave only references in the bibliography and move all
%% accessory text in footnotes.
%% (iii) Also, please have only one work for each \bibitem.

\end{document}